\documentclass[11pt,showpacs]{article}
\pdfoutput=1
\usepackage{color}
\usepackage[utf8]{inputenc}
\usepackage{caption}
\usepackage{subcaption}
\usepackage{amsmath}
\usepackage{amsfonts}
\usepackage{cite}
\usepackage{amssymb}
\usepackage{graphicx}
\usepackage{dcolumn}
\usepackage{bm}
\usepackage{proof}
\usepackage{multirow}
\usepackage[figuresright]{rotating}
\usepackage[left=2cm,right=2cm,top=2cm,bottom=2cm]{geometry}

\begin{document}


\title{Linear instability of channel flow with microgroove-type anisotropic superhydrophobic walls}

\author{Xueyan Zhai$^*$,  Kaiwen Chen\footnote{Xueyan Zhai and Kaiwen Chen contributed equally to this work.} , Baofang Song
\thanks{Email address for correspondence: baofang\_song@tju.edu.cn} \\
\small{Center for Applied Mathematics, Tianjin University, Tianjin 300072, China}
}


\maketitle

\begin{abstract}
We study the temporal linear instability of channel flow subject to a tensorial slip boundary condition that models the slip effect induced by microgroove-type super-hydrophobic surfaces. 
The microgrooves are not necessarily aligned with the driving pressure gradient.
Pralits {\it et al.} Phys. Rev. Fluids $\bm 2$, 013901 (2017) investigated the same problem and reported that a proper tilt angle of the microgrooves about the driving pressure gradient can reduce the critical Reynolds number and that the flow with a single superhydrophobic wall is much more unstable/less stable than that with two superhydrophobic walls. In contrast, we show that the lowest critical Reynolds number is always realized with two superhydrophobic walls, and we obtain critical Reynolds numbers significantly lower than the reported. 
Besides, we show that the critical Reynolds number can be further reduced by increasing the anisotropy in the slip length. \textcolor{black}{As the tilt angle changes, there appears to be a strong correlation between the strength of the instability and the magnitude of the cross-flow component of the base flow incurred by the tilt angle.} In case the tilt angles of the microgrooves differ on the two walls, the critical Reynolds number increases as the difference in the tilt angles increases, i.e. two superhydrophobic walls with parallel microgrooves give the lowest critical Reynolds number.
The results are informative for designing the microgroove-type wall texture to introduce instability at low Reynolds number channel flow, which may be of interest for enhancing mixing or heat transfer in small flow systems where turbulence cannot be triggered. 
\end{abstract}

\section{Introduction}{\label{sec:intro}}
Significant velocity slip can be obtained by properly texturing the wall surfaces in viscous flows \cite{Qu2005, Choi2006, Lee2008, Lee2009, Voronov2008, Rothstein2010, Chattopadhyay2018}.
Slip boundary condition on a smooth wall is a simplified treatment of complex superhydrophobic surfaces. This simplification has been applied to various flow problems ranging from linear stability analysis to flow transition and even turbulent flows \cite{Chu2004, Lauga2005, Min2005, Ghosh2014a, Ghosh2014b, Seo2016, Yu2016, Chai2019, Picella2019, Picella2020, Xiong2020, Pralits2017, Davis2020, Gersting1974,Jouin2022}. \textcolor{black}{Among these, Refs. \cite{Yu2016, Picella2020, Seo2016} particularly showed the applicability of this simplification to linear stability, transition and even turbulence problems.} 
Regarding the linear stability analysis of slip channel flow, a few studies had concluded that velocity slip stabilizes the flow and greatly increases the critical Reynolds number \cite{Lauga2005,Min2005,Gersting1974}. 

However, a few recent studies reported that velocity slip does not always stabilize the flow, but can also destabilize the flow given a certain amount of anisotropy in the slip \cite{Chai2019, Xiong2020, Pralits2017, Chen2021,Jouin2022}. In fact, the significant stabilizing effect concluded before was clarified to be only for two-dimensional (2-D) perturbations. Three-dimensional (3-D) modes become dominant if sufficiently strong anisotropy in the slip length (the usual measure of the slip effect) is taken into consideration, and can be destabilized by the slip at Reynolds numbers far below the critical Reynolds number for 2-D modes. Among these studies, Refs. \cite{Chai2019} and \cite{Xiong2020} considered a special case where the slip in streamwise and spanwise directions are independent of each other, i.e.
\begin{equation}\label{equ:BC}
\lambda_x\frac{\partial u}{\partial n}+u=0, \hspace{3mm}
\lambda_z\frac{\partial w}{\partial n}+w=0,
\end{equation}
at the channel wall, where $x, y, z$ denote the coordinates in the direction of the driving pressure gradient (will be simply referred to as the streamwise direction hereafter), wall normal and spanwise directions, respectively, $u$ and $w$ the streamwise and spanwise velocities, and $\lambda_x$ and $\lambda_z$ the slip lengths associated with $u$ and $w$, respectively. In the cases of pure streamwise slip and pure spanwise slip, as considered by \cite{Chai2019}, it was shown that the leading mode becomes 3-D when $\lambda_x$ is larger than approximately 0.008 and when $\lambda_z$ is larger than 0.02, respectively. Streamwise slip only modestly increases the critical Reynolds number compared to the no-slip case, whereas spanwise slip can greatly reduce the critical Reynolds number to a few hundred when $\lambda_z$ is increased to above 0.1. 
Ref. \cite{Xiong2020} conducted a thorough theoretical study when slip is present in both directions and confirmed the findings of \cite{Chai2019}. Their asymptotic analysis also gives the analytic dependence of the critical Reynolds number on the slip length in the small slip length regime. 

In fact, the boundary condition (\ref{equ:BC}) is highly idealized. For superhydrophobic surfaces with complex textures, more realistically, the boundary condition should take the tensorial form \cite{Vinogradova1999,Asmolov2012,Pralits2017}. For parallel-microgroove-type textures, which is considered in this paper, Ref. \cite{Pralits2017} presented the boundary condition
\begin{equation} \label{equ:BC2}
\left[\begin{array}{c}
u \\ 
w 
\end{array}\right]
+\Lambda\frac{\partial}{\partial n}
\left[\begin{array}{c}
u \\ 
w
\end{array}\right]
=0
\end{equation} 
at the top and bottom walls, where $n$ denotes the outward normal direction at the wall. The slip tensor $\Lambda$ takes the following form
\begin{equation}
\bm \Lambda=\bm Q
\left[\begin{array}{c c}
\lambda^\parallel & 0 \\
0 & \lambda^\bot
\end{array}\right]
\bm Q^T, \text{  with  } 
\bm Q=
\left[\begin{array}{c c}
\cos{\theta} & -\sin{\theta} \\
\sin{\theta} & \cos{\theta}
\end{array}\right],
\end{equation}
where $\lambda^\parallel$ and $\lambda^\bot$ are the eigenvalues of the slip tensor $\Lambda$, corresponding to the longitudinal (parallel to the grooves) and transverse (perpendicular to the grooves) slip lengths, respectively,
and $\theta$ is the angle of the alignment of the microgrooves about the streamwise direction. When $\theta$ differs from $0^\circ$ and $90^\circ$, $u$ and $w$ will be coupled with each other through the boundary condition (\ref{equ:BC2}). The boundary condition (\ref{equ:BC}) used in \cite{Chai2019, Xiong2020} is actually the special case of (\ref{equ:BC2}) for $\theta=0^\circ$ and $90^\circ$, i.e. when the microgrooves are parallel with or perpendicular to the driving pressure gradient, as suggested by Refs. \cite{Gogte2005, Belyaev2010}. With boundary condition (\ref{equ:BC2}), Ref. \cite{Pralits2017} investigated the dependence of the critical Reynolds number on the slip length and tilt angle $\theta$ thoroughly. The results showed that, with a non-vanishing tilt angle, the leading mode also becomes 3-D and the critical Reynolds number can be reduced compared to the no-slip case, and the authors also discussed about why Squire's theorem does not necessarily hold in the case of anisotropic slip. It was also reported that the destabilizing effect is much more prominent in the case with a single superhydrophobic channel wall, and the destabilizing effect seemed to maximize at a tilt angle close to $\theta=45^\circ$ for either a single or two superhydrophobic walls. \textcolor{black}{Ref. \cite{Jouin2022} investigated the linear stability and transition problem of channel flow subject to this type of boundary condition by fixing $\theta=45^\circ$ and confirmed the destabilizing effect of the anisotropic slip.}  
\textcolor{black}{By considering the actually alternate solid/gas configuration on microgrooved surfaces (modelled as no-slip/full slip regions) instead of a smooth slippery wall with homogeneous slip lengths, Ref.~\cite{Yu2016} and \cite{Tomlinson2022} both reported destabilizing effects of the microgrooves even with a zero tilt angle using a bi-global stability analysis, though  didn't perform a parametric study of the effect of the anisotropy of the slip on the instability.}
All these studies suggest the possibility of introducing instability or earlier transition in flow problems at lower Reynolds numbers (compared to the no-slip case) by using properly textured superhydrophobic surfaces.

In this paper, linear stability of channel flow subject to the tensorial boundary condition \eqref{equ:BC2} is considered and the aim is threefold. Firstly, to verify whether or not a single superhydrophobic wall indeed results in much lower critical Reynolds number than two superhydrophobic walls, as reported by \cite{Pralits2017}. %
Secondly, Ref. \cite{Pralits2017} only considered $\lambda^\parallel/\lambda^\bot=2$ in their analysis,
while we consider a larger range of this ratio up to 10 to investigate the possibility of lower critical Reynolds numbers than the reported. This is because some studies have suggested larger values of this ratio \cite{Ng2009}. Thirdly, we consider the stability in the presence of two superhydrophobic walls where the microgrooves are non-parallel on the two walls. We believe the results are informative for enhancing mixing, heat transfer and flow control by designing the wall texture in small channel flow systems, where turbulence cannot be triggered due to the low Reynolds number. 

\section{Methods}\label{sec:methods}

We consider the nondimensional incompressible Navier-Stokes equations
\begin{equation}\label{equ:NS}
 \frac{\partial \bm u}{\partial t}+{\bm u}\cdot\bm{\nabla}
{\bm u}=-{\bm{\nabla}p}+\frac{1}{Re}{\bm\nabla^2}{\bm u}, \;
\bm{\nabla}\cdot{\bm u}=0
\end{equation}
for a channel flow in Cartesian coordinates $(x, y, z)$, where $\bm u=(u,v,w)$ denotes velocity vector, $p$ denotes pressure, respectively. For comparison with \cite{Pralits2017}, velocities are normalized by $U_b$, i.e. the average streamwise velocity on the $z-y$ channel cross-section, length by half gap width $h$ and time by $h/U_b$, and therefore the Reynolds number $Re=U_bh/\nu$ where $\nu$ is the kinematic viscosity of the fluid. The origin of the $y$-axis is placed at the channel center. 

\subsection{The linearized equations}\label{sec:linearization}

Here we denote the fully developed base flow as 
\begin{equation}
{\bm U}=U(y){\bm e}_x + W(y){\bm e}_z,
\end{equation} where ${\bm e}_x$ and $\bm e_z$ are the unit vectors in the streamwise and spanwise directions, respectively. 
In the case of two superhydrophobic walls with identical characteristics, i.e. $\theta$ and $\lambda^{\parallel}/\lambda^{\bot}$ are respectively identical on the two walls, the basic flow is
\begin{equation}
	U(y)=1-\dfrac{y^2-1/3}{2\lambda^\parallel \cos^2\theta+2\lambda^\bot \sin^2\theta+2/3},
\end{equation}
\begin{equation}
	W(y)=\dfrac{(\lambda^\parallel-\lambda^\bot)\cos\theta \sin\theta}{\lambda^\parallel \cos^2\theta+\lambda^\bot \sin^2\theta+1/3},
\end{equation}
where the spanwise component $W$ is a constant.
In the case where only the bottom wall is superhydrophobic, the basic flow is
\begin{equation}
      U(y)=-\dfrac{y^2-1+(2l^\parallel \cos^2\theta+2l^\bot \sin^2\theta)(y-1)}{2/3+2l^\parallel \cos^2\theta+2l^\bot \sin^2\theta},
\end{equation}
and
\begin{equation}
	W(y)=-\dfrac{2(l^\parallel-l^\bot)\cos\theta \sin\theta(y-1)}{2/3+2l^\parallel \cos^2\theta+2l^\bot \sin^2\theta},
\end{equation}
where
\begin{equation}
	l^\parallel=\dfrac{\lambda^\parallel}{2+\lambda^\parallel},\hspace{3mm} l^\bot=\dfrac{\lambda^\bot}{2+\lambda^\bot},
\end{equation}
and the spanwise component $W$ is a linear function of $y$. \textcolor{black}{Note that the volume flux associated with the streamwise component $U$ is fixed following Ref.~\cite{Pralits2017}.}

\begin{figure}
\centering
\includegraphics[width=0.99\linewidth]{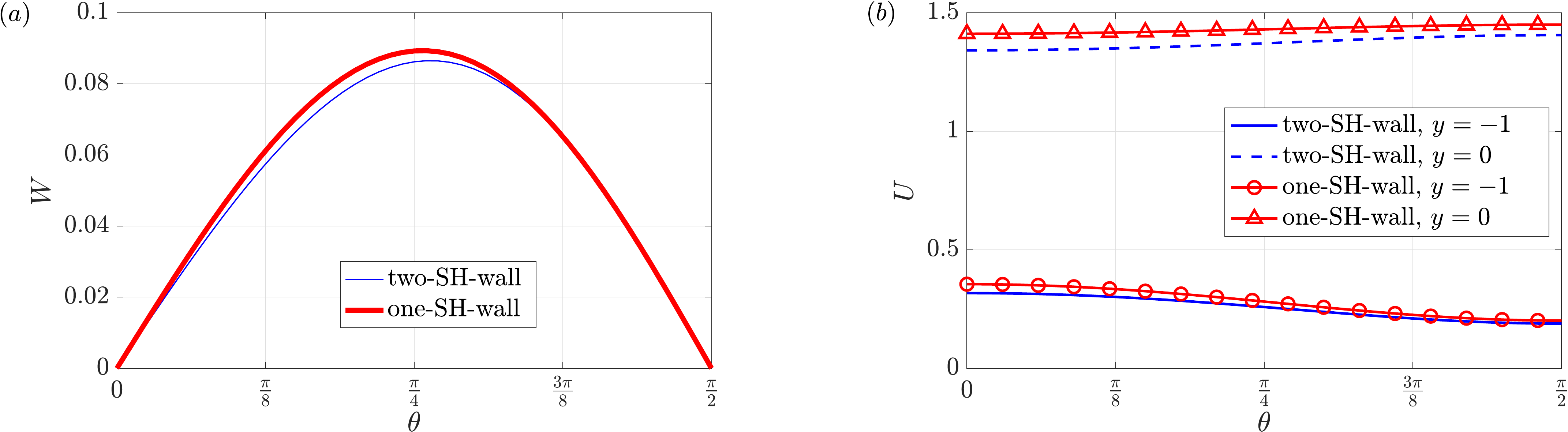}
\caption[wave number]{
    \label{fig:baseflow} 
     The effect of the tilt angle $\theta$ on the base flow. (a) The maximum of $W(y)$ over $y$. Note that for the one-SH-wall case, $W(y)$ maximizes at the slippery wall at $y=-1$. (b) The streamwise velocity at the bottom slippery wall $U(y=-1)$ and at the channel center $U(y=0)$. The slip parameters are $\lambda^\parallel=0.155$ and $\lambda^\bot=\lambda^\parallel/2$.
} 
\end{figure}

\textcolor{black}{Obviously, a tilt angle differing from 0 and $\pi/2$ will result in a non-vanishing spanwise velocity component in the base flow, i.e. a cross-flow component. Figure~\ref{fig:baseflow} shows the effect of the tilt angle on the base flow for $\lambda^\parallel=0.155$ and $\lambda^\bot=\lambda^\parallel/2$. It can be seen that the cross-flow component is maximized slightly above $\theta=\pi/4$. The streamwise velocity at the channel center increases whereas the slip velocity at the wall decreases with the tilt angle. The trends for $U$ are expected given that $\lambda^\parallel > \lambda^\bot$ and its volume flux is fixed.}

Note that, although we give the analytic form of the basic flows for the two specific cases, the form may be much more complicated in more general cases, e.g. in cases where $\theta$ or the ratio $\lambda^{\parallel}/\lambda^{\bot}$ takes different non-vanishing values on the two walls. It would be much easier to solve for the basic flow numerically from the governing equations in these situations, and the numerical solution can also be used for the linear stability analysis.  

Introducing small disturbances $\bm u=(u,v,w)$ and linearizing the Navier-Stokes equations about the base flow, we obtain the governing equation for $\bm u$ as the following,

\begin{equation}\label{equ:LNS}
 \frac{\partial \bm u}{\partial t}+\bm u\cdot\bm\nabla \bm U 
+\bm U\cdot\bm{\nabla}{\bm u} = -\bm\nabla p+\frac{1}{Re}{\bm\nabla}^2 \bm u, \hspace{3mm}
\nabla \cdot \bm u  = 0
\end{equation}
with the condition Eqs.~\eqref{equ:BC2} and impermeability condition for $\bm u$. In the following, we introduce three different formulations for the eigenvalue analysis, i.e. primitive variable formulation, velocity-vorticity formulation and direct simulation of the Navier-Stokes equations.
\subsection{Primitive variable formulation ($u-p$ formulation)}
A disturbance is expressed in terms of Fourier modes along the wall-parallel directions,
\begin{equation}\label{equ:distur}
	q(x,y,z,t)=\hat{q}(y) e^{i(\alpha x+\beta z-\omega t)}+c.c.,
\end{equation}
where $\alpha$ and $\beta$ are the streamwise and spanwise wavenumbers, respectively, $\hat q$ is the Fourier coefficient, and c.c. denotes the complex conjugate. The complex angular frequency is denoted as $ \omega = \omega_r + i\omega_i$ (subscript $i$ denotes the imaginary part, subscript $r$ denotes the real part) and $\omega_i > 0$ indicates a linear instability.  Plugging into Eqs. \eqref{equ:LNS}, we get 
\begin{equation}
	-i\omega \bm {\hat{q}}=\bm L \bm {\hat{q}}.
\end{equation}
This is an eigenvalue problem, where

\begin{equation}
	\bm L=
	\left[ \begin{array}{cccc}
		A & -\frac{\partial U}{\partial y} & 0 &  -i \alpha \\
		0 & A  & 0 &  -\frac{\partial }{\partial y}       \\
		0 & -\frac{\partial W}{\partial y} & A & -i\beta \\
		i \alpha & \frac{\partial }{\partial y} &i\beta & 0 \\
	\end{array} 
	\right ],\hspace{3mm}
	\bm {\hat{q}}=
		\left[ \begin{array}{cccc}
		\hat{u}  \\
		\hat{v}  \\
		\hat{w}  \\
		\hat{p}  \\
	\end{array}
	\right ],
\end{equation}

\begin{equation}
	A=\dfrac{1}{Re}(\dfrac{\partial^2}{\partial y^2}-\alpha^2-\beta^2)-i\alpha U-i\beta W.
\end{equation}
We use a Chebyshev-collocation discretization in the wall-normal direction \cite{Trefethen2000}. The operator $\bm L$ is a $4N \times 4N$ complex matrix after discretization, where $N$ is the number of grid points taken in the $y$ direction.

\subsection{Velocity-Vorticity formulation ($v-\eta$ formulation)}
We also consider the velocity-vorticity formulation of the linearized Navier-Stokes equations \cite{Reddy1993} for cross-validation. 
The linearized equations in this formulation read
\begin{eqnarray} \label{equ:veta}
 \left(\dfrac{\partial}{\partial t}+U\dfrac{\partial}{\partial x}+W\dfrac{\partial}{\partial z}\right)\nabla^2v - \dfrac{d^2U}{dy^2}\dfrac{\partial v}{\partial x} - \dfrac{d^2W}{dy^2}\dfrac{\partial v}{\partial z} & =\dfrac{1}{Re}\nabla^4v, \\
\label{equ:veta2}
 \left(\dfrac{\partial}{\partial t}+U\dfrac{\partial}{\partial x}+W\dfrac{\partial}{\partial z}\right)\eta + \dfrac{dU}{dy}\dfrac{\partial v}{\partial z} - \dfrac{dW}{dy}\dfrac{\partial v}{\partial x} &=\dfrac{1}{Re}\nabla^2\eta,
\end{eqnarray}
where $\eta=\partial u/\partial z-\partial w/\partial x$ is the $y$-component of the vorticity. 
Using the incompressibility condition, $u$ and $w$ can be derived in spectral space as
\begin{eqnarray}
\hat u &=\dfrac{1}{i(\alpha^2+\beta^2)}\left(\beta \hat\eta -\alpha\dfrac{\partial\hat v}{\partial y}\right),\\
\hat w &=\dfrac{1}{i(\alpha^2+\beta^2)}\left(-\alpha\hat\eta-\beta \dfrac{\partial\hat v}{\partial y}\right).
\end{eqnarray}
Further, the boundary condition for $\eta$ can be derived using the slip boundary condition \eqref{equ:BC2}. The same Chebyshev-collocation discretization for the primitive variable formulation is used here for discretizing the linear operators.
Substituting into Eqs. \eqref{equ:veta} and \eqref{equ:veta2}, we get an eigenvalue problem
\begin{equation}
	-i\omega \bm {\hat{q}}=\bm L \bm {\hat{q}},
\end{equation}
where

\begin{equation}
	\bm L=
	-i\left[ \begin{array}{cc}
		\bm L_{os} & 0 \\
		\bm L_c & \bm L_{sq}         \\
	\end{array} 
	\right ],\hspace{3mm}
	\bm {\hat{q}}=
	\left[ \begin{array}{cc}
		\hat{v}  \\
		\hat{\eta}  \\
	\end{array} 
	\right ],
\end{equation}
\begin{equation}
	\bm L_{os}=-D_k^{-1}[D_k^2/(iRe)-\alpha UD_k+\alpha D^2U-\beta WD_k+\beta D^2W],
\end{equation}
\begin{equation}
	\bm L_{c}=\beta DU-\alpha DW,
\end{equation}
\begin{equation}
	\bm L_{sq}=\alpha U+\beta W-D_k/(iRe),
\end{equation}
\begin{equation}
	D_k=D^2-k^2,\hspace{3mm} k^2=\alpha^2+\beta^2.
\end{equation}
The operator $\bm L$ is a complex $2N \times 2N$ matrix, where $N$ is the number of grid points taken in the $y$ direction.

\subsection{Time-stepping the Navier-Stokes equations (DNS formulation)}
For further validation of our eigenvalue calculations, we also solve the \textcolor{black}{linearized} Navier-Stokes equations (\ref{equ:LNS}) using a Fourier-spectral-finite-difference scheme. Periodic boundary conditions are imposed and Fourier spectral method is used for the spatial discretization in the streamwise and spanwise directions. In the wall normal direction, a Chebyshev collocation method \cite{Trefethen2000} is used for the spatial discretization. For the channel geometry, the ($\alpha, \beta)$ mode of velocity and pressure field is expressed as Eqs. \eqref{equ:distur}.
The integration in time is performed using a second-order-accurate Adams-Bashforth/backward differentiation scheme \cite{Hugues1998}. The incompressibility condition and boundary condition (\ref{equ:BC2}) are imposed using the influence matrix technique (\cite{Phillips1991,Willis2017}, see details in Appendix \ref{sec:NS_solver}). 
The largest eigenvalue of a given Fourier mode can be calculated from the time-series of the amplitude of the velocity disturbances or modal kinetic energy when the leading eigenmode has become dominant in the flow field.

\textcolor{black}{The convergence test regarding the grid resolution and the cross validation of the three formulations are shown in Appendix~\ref{sec:validation}.}

\section{Results}

\subsection{Distribution of eigenvalue in the wavenumber plane}

\begin{figure}
\centering
\includegraphics[width=0.99\linewidth]{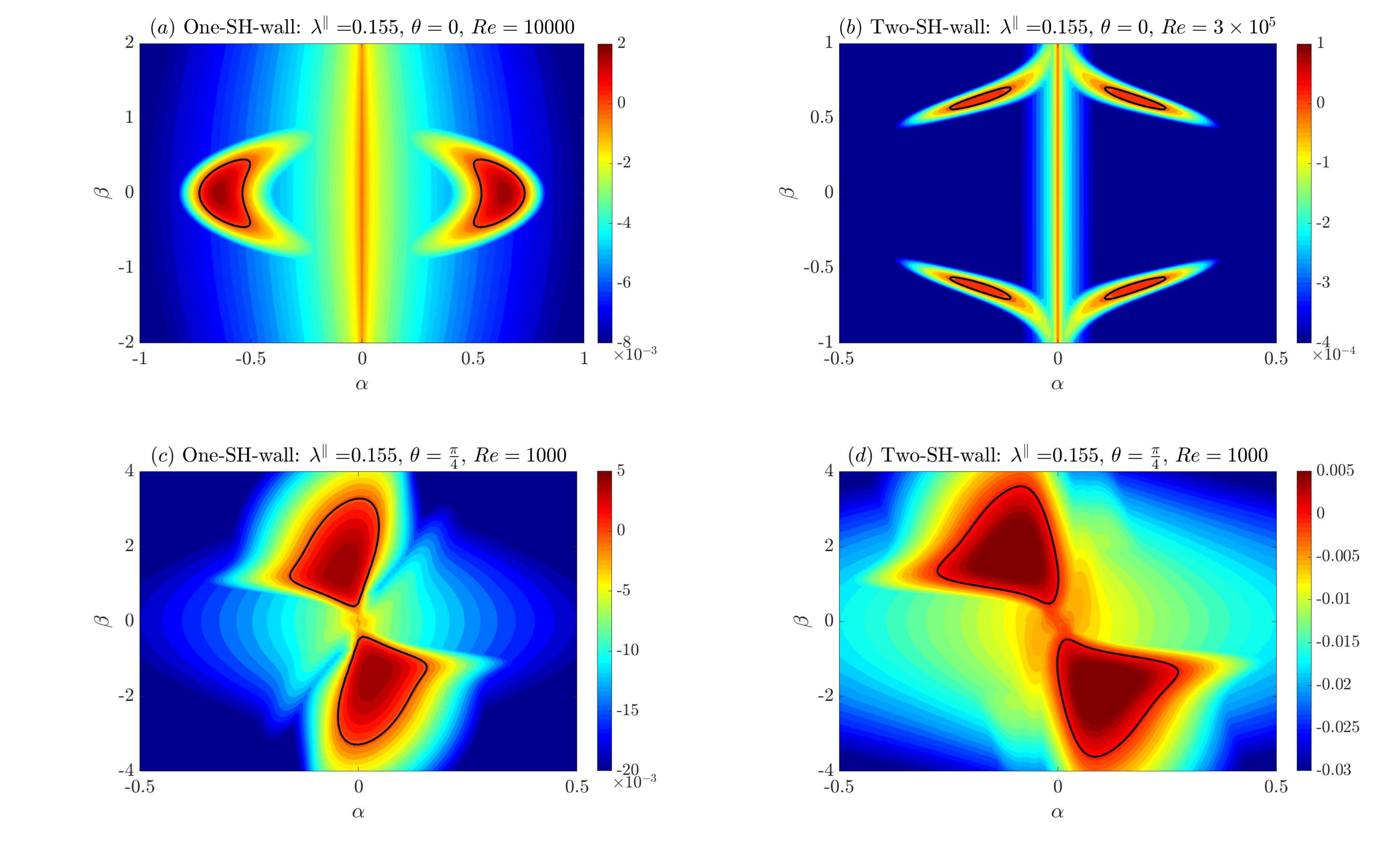}
\caption[wavenumber plane]{
    \label{fig:unstable_region} 
    The effects of the tilt angle on the distribution of eigenvalue $\omega_i$ in the wavenumber plane. Slip lengths are set to be $\lambda^\parallel=0.155$ and $\lambda^\bot=\lambda^\parallel/2$. (a) One-SH-wall with $\theta=0$ and $Re=10000$. (b) Two-SH-wall with $\theta=0$ and $Re=3\times 10^5$. (c) One-SH-wall with $\theta=\pi/4$ and $Re=1000$. (d) Two-SH-wall with $\theta=\pi/4$ and $Re=1000$. The bold line encloses the linearly unstable region in the wavenumber plane.
} 
\end{figure}

\textcolor{black}{In the no-slip case, the distribution of eigenvalue ($\omega_i$) is symmetric about $\beta=0$, i.e. one only needs to search in the first quadrant of the $\alpha-\beta$ wavenumber plane for unstable modes. Besides, the leading mode is 2-D according to Squire's theorem so that practically one only needs to search the non-negative part of the $\alpha-$axis with $\beta=0$ for the leading mode. However, when there is a tilt angle $\theta$ of the microgrooves, a cross-flow component of the base flow appears and the symmetry of the base flow is broken. Therefore, the distribution of the eigenvalue in the $\alpha-\beta$ plane should be changed by the tilt angle also, see an example in figure~\ref{fig:unstable_region}. If $\theta=0$, clearly the cross-flow component $W$ is zero, therefore, the distribution of $\omega_i$ appears to be symmetric about $\beta=0$ as in the no-slip case, see figure~\ref{fig:unstable_region}(a, b). The difference between the two slip settings is that the most unstable mode of the one-SH-wall case appears to be 2-D with $\beta=0$, whereas to be 3-D with a non-zero $\beta$ in the two-SH-wall case. Similarly, the distribution of eigenvalue is also symmetric about $\beta=0$ if $\theta=\pi/2$ because the cross-flow component $W$ is zero also. If $\theta=\pi/4$, the cross-flow $W$ is non-zero, and figure~\ref{fig:unstable_region}(c, d) shows that the symmetry in the distribution of $\omega_i$ about $\beta=0$ is broken. If considering $\alpha\geq 0$, the unstable region appears to be mainly located in the negative-$\beta$ region, i.e. the fourth quadrant in the wavenumber plane, and the flow is mainly stable in the first quadrant. Note that the sign of the wavenumber determines the orientation of the wave structure. Therefore, to search for unstable modes, one has to scan through the whole right half-plane instead of only the first quadrant. Ref. \cite{Jouin2022} also shows this symmetry-breaking effect by the cross-flow resulting from the tilt angle.
In fact, this effect is also present in other problems with cross-flow-related instabilities, such as 3-D boundary layer flow over swept wings \cite{Mack1984,Saric2003} and localized turbulent bands in channel flow \cite{Xiao2020,Song2020} under no-slip boundary condition. The remaining part of this paper focuses on the critical Reynolds number $Re_{cr}$, which can be obtained by searching for the first appearance of an unstable mode in the right half wavenumber plane as $Re$ increases.}

\subsection{Dependence of the critical Reynolds number on the angle $\theta$}
\label{sec:critical_Re_vs_angle}

\begin{figure}
\centering
\includegraphics[width=0.9\linewidth]{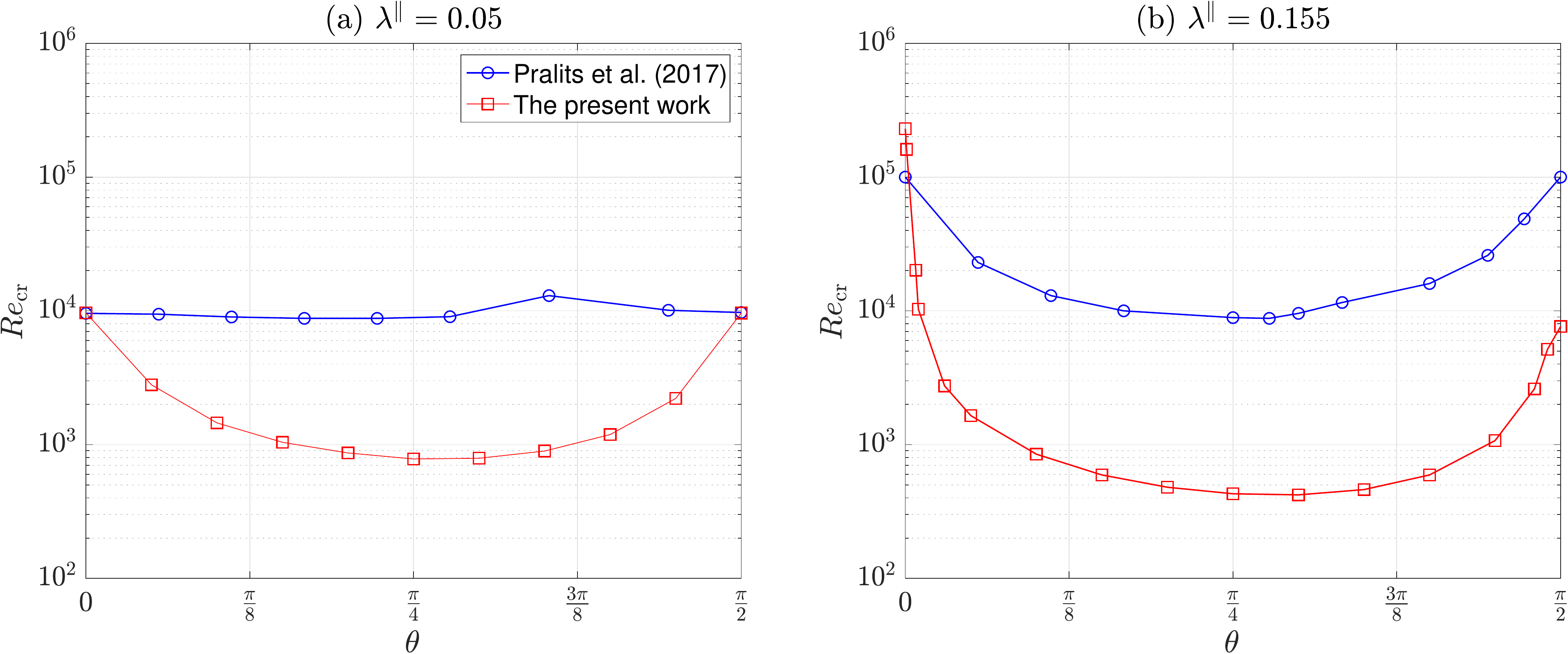}
\caption[validation]{
    \label{fig:critical_Re_vs_angle} 
     The critical Reynolds number as a function of the tilt angle $\theta$ for the two-SH-wall channel. (a) $\lambda^\parallel=0.05$; (b) $\lambda^\parallel=0.155$. Our calculations are shown as red squares and the results of \cite{Pralits2017} are shown as blue circles for comparison.
} 
\end{figure}

In this section, we set $\lambda^\parallel/\lambda^\bot=2$ as Ref. \cite{Pralits2017} and investigate the dependence of the critical Reynolds number on the tilt angle $\theta$ of the microgrooves. All the calculations are performed using the $u-p$ formulation. Unless explicitly specified, the number of wall-normal grid points is set to 128 (half and double resolutions give nearly the same leading eigenvalue in the parameter regime considered here, see table~\ref{tab:convergence} in Appendix \ref{sec:validation}).

\begin{figure}
\centering
\includegraphics[width=0.99\linewidth]{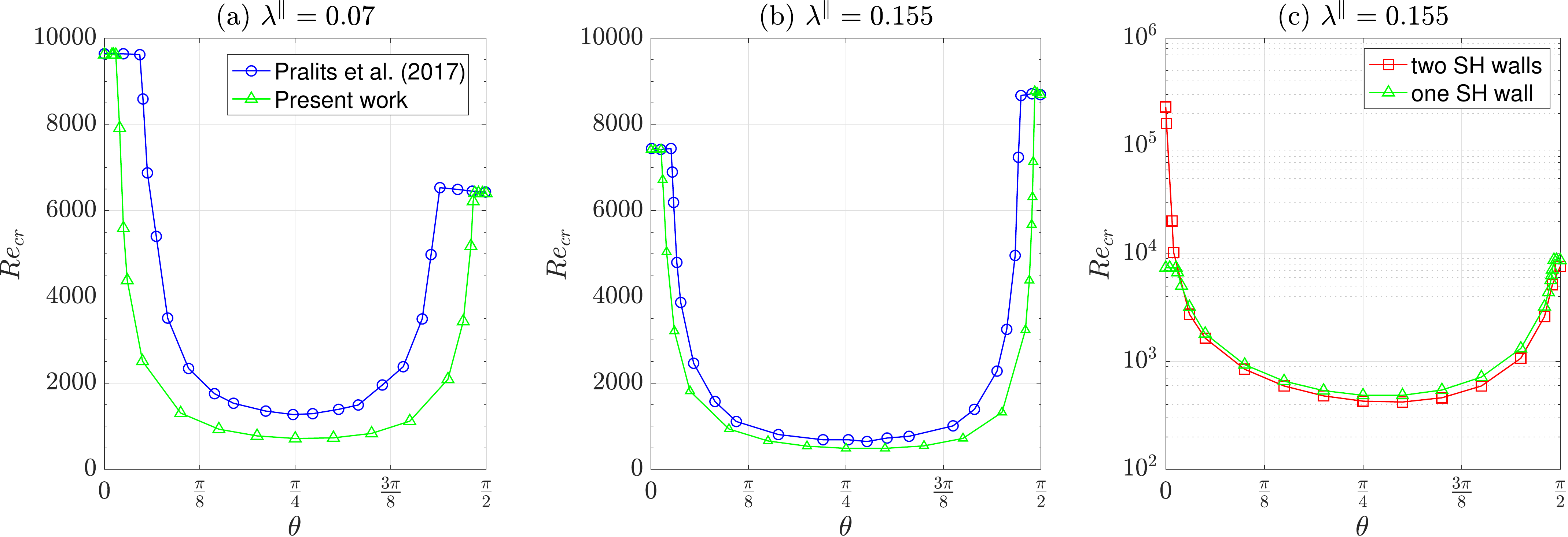}
\caption[validation]{
    \label{fig:critical_Re_vs_angle_one_wall} 
     (a,b) The critical Reynolds number as a function of the tilt angle $\theta$ for the one-SH-wall channel. Our calculations are shown as green triangles and the results of \cite{Pralits2017} are shown as blue circles for comparison. (c) The critical Reynolds number for two-SH-wall and one-SH-wall cases in the present work are compared. 
} 
\end{figure}

\begin{figure}
\centering
\includegraphics[width=0.99\linewidth]{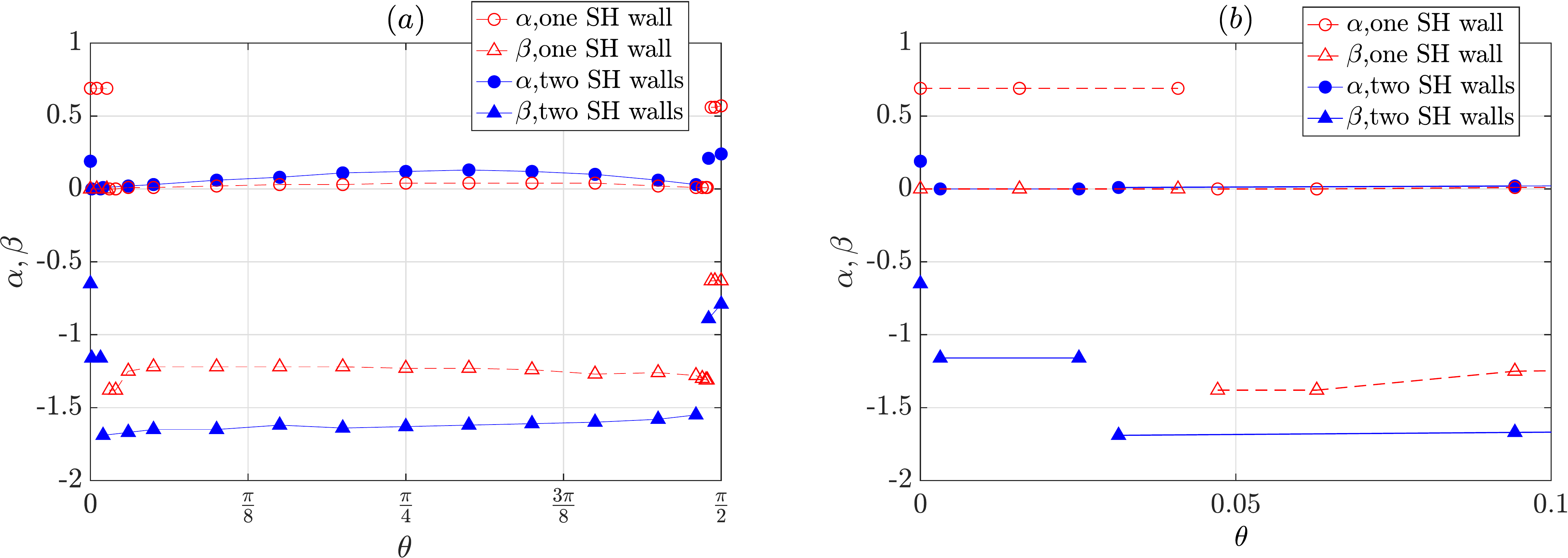}
\caption[wave number]{
    \label{fig:most_unstable_wave_number} 
     (a) The most unstable wavenumbers at the critical Reynolds number as functions of tilt angle $\theta$ for two-SH-wall and one-SH-wall cases. (b) A close-up at small $\theta$.
} 
\end{figure}

Figure \ref{fig:critical_Re_vs_angle} shows the critical Reynolds number as a function of the angle $\theta$ for $\lambda^\parallel=0.05$ and $\lambda^\parallel=0.155$ in the two-SH-wall channel. The results of \cite{Pralits2017} are also shown for comparison, which disagree with our calculations for both slip lengths. The critical Reynolds numbers calculated here are much lower than those of \cite{Pralits2017} for nearly all $\theta$ values. The largest difference is around $\theta=\pi/4$ and is as large as an order of magnitude in a wide range of $\theta$, especially for the $\lambda^\parallel=0.155$ case. Interestingly, for the $\lambda^\parallel=0.05$ case, our critical Reynolds numbers agree with that of \cite{Pralits2017} for $\theta=0^\circ$ and $90^\circ$. In the $\lambda^\parallel=0.155$ case, our critical Reynold number is higher than that of \cite{Pralits2017} at $\theta=0^\circ$, whereas is lower by more than one order of magnitude at $\theta=90^\circ$.

Similar calculations are performed for the one-SH-wall channel also. Figure \ref{fig:critical_Re_vs_angle_one_wall} shows the critical Reynolds number for $\lambda^\parallel=0.07$ and 0.155. The disagreement between our results and Ref. \cite{Pralits2017} is still noticeable, although smaller than the disagreement for the two-SH-wall channel. Our results give lower critical Reynolds numbers for nearly all $\theta$ values. Nevertheless, the trends of $Re_{cr}$ as $\theta$ changes in both studies are similar, exhibiting a U-shape minimizing close to $\theta=\pi/4$. In figure \ref{fig:critical_Re_vs_angle_one_wall}(c), the $Re_{cr}$ for two-SH-wall and one-SH-wall channels with $\lambda^{\parallel}=0.155$ are compared. It can be seen that $Re_{cr}$ is slightly lower in the two-SH-wall case than in the one-SH-wall case, except for at vanishing $\theta$, where the one-SH-wall flow is much more unstable than the two-SH-wall flow. This observation disagrees with the conclusion of \cite{Pralits2017} that the one-SH-wall case is much more unstable than the two-SH-wall case for all $\theta$ values under the same slip parameters. 

\begin{figure}
\centering
\includegraphics[width=0.99\linewidth]{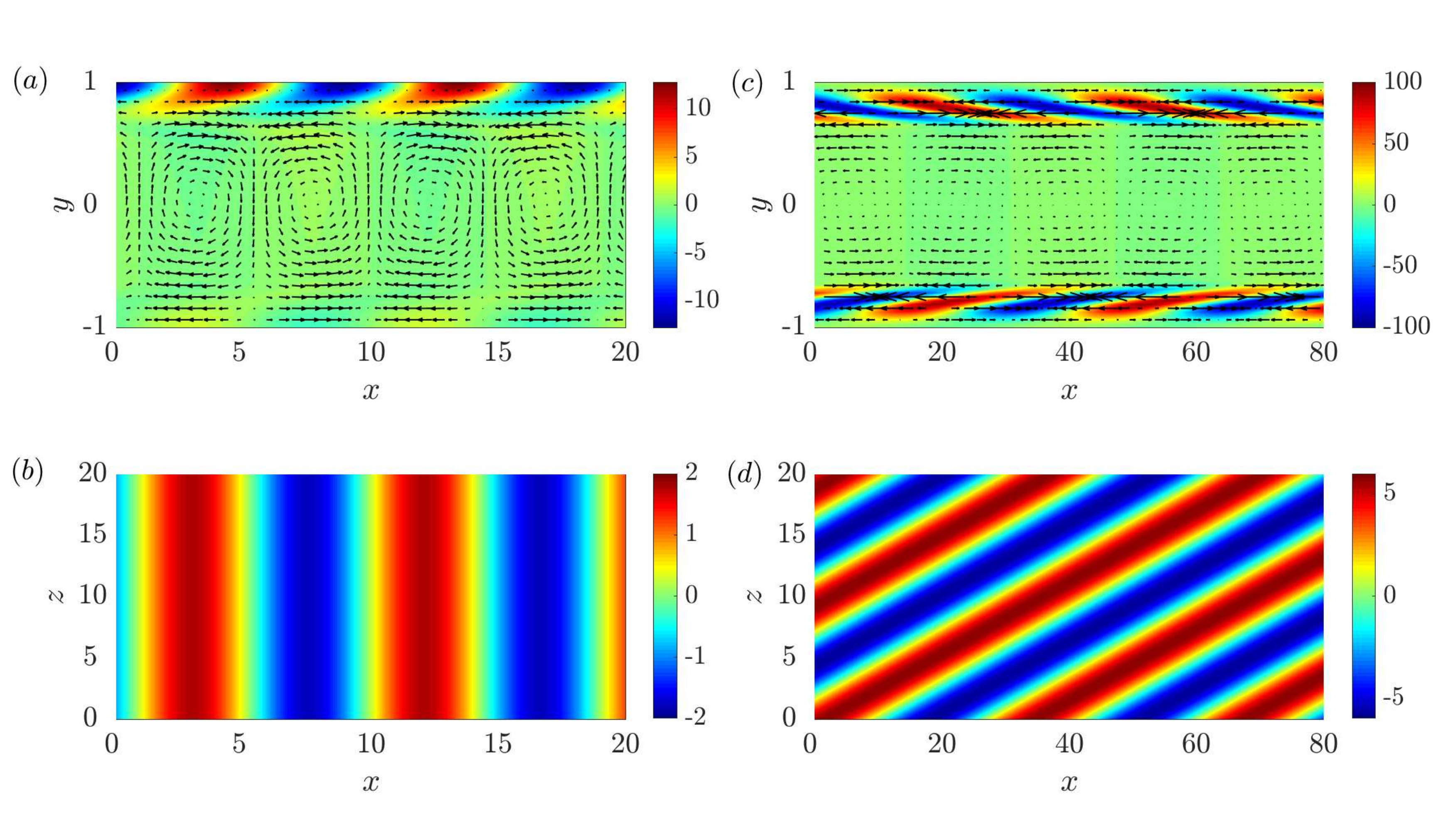}
\caption[wave number]{
    \label{fig:flow_structure_0degree} 
     The flow structure of the leading eigenmode for the one-SH-wall $(a, b)$ and two-SH-wall $(c, d)$ cases with $\theta=0$ at the respective critical Reynolds number. In $(a, b)$, the bottom wall is slippery, the Reynolds number is $Re=7417$ and the wavenumbers are $(\alpha, \beta)=(0.69, 0)$. In $(c, d)$, the Reynolds number is $Re=2.3\times 10^5$ and the wavenumbers are $(\alpha, \beta)=(0.19, -0.65)$. The color shows the spanwise vorticity on the bottom wall and vectors show the in-plane velocities. The magnitudes of the shown quantities are arbitrary.
} 
\end{figure}

In order to understand why the critical Reynolds number differs largely close to $\theta=0$ for the one-SH-wall and two-SH-wall cases, the critical wavenumbers as functions of $\theta$ for $\lambda^{\parallel}=0.155$ are shown in figure \ref{fig:most_unstable_wave_number}. It can be seen that, the critical wavenumbers experience complicated and sharp changes in the small $\theta$ regime. In the one-SH-wall case, the streamwise wavenumber stays constant at about $\alpha_c=0.7$ before suddenly drops to $\alpha_c\approx 0$ as $\theta$ is increased to approximately 0.04, and the spanwise wavenumber stays constant at $\beta_c=0$ (see figure \ref{fig:unstable_region}a) before suddenly transitions to about $\beta_c=-1.2$ 
 This change indicates that the most unstable perturbation transitions sharply from a spanwise-invariant ($\beta=0$) one to a nearly streamwise-invariant ($\alpha\approx 0$) one as $\theta$ is increased in the small $\theta$ regime. In contrast, in the two-SH-wall case, the most unstable perturbation is 3-D with non-vanishing $\alpha_c$ and $\beta_c$ at $\theta=0$. However, immediately above $\theta=0$, the most unstable perturbation transitions to a two dimensional streamwise-invariant ($\alpha=0$) one, and then transitions to a nearly streamwise-invariant ones with extremely small $\alpha_c$ at about $\theta=0.03$ ($\approx 1.7^\circ$). The significant difference in the critical Reynolds number for $\theta=0$ is attributed to the distinct flow characteristics in the small $\theta$ regime. For larger $\theta$ ($\geqslant 0.04$ or $2^\circ$), the most unstable perturbations are always three dimensional with non-vanishing $\alpha$ and $\beta$ and the trends as $\theta$ changes in both cases are similar.

\begin{figure}
\centering
\includegraphics[width=0.99\linewidth]{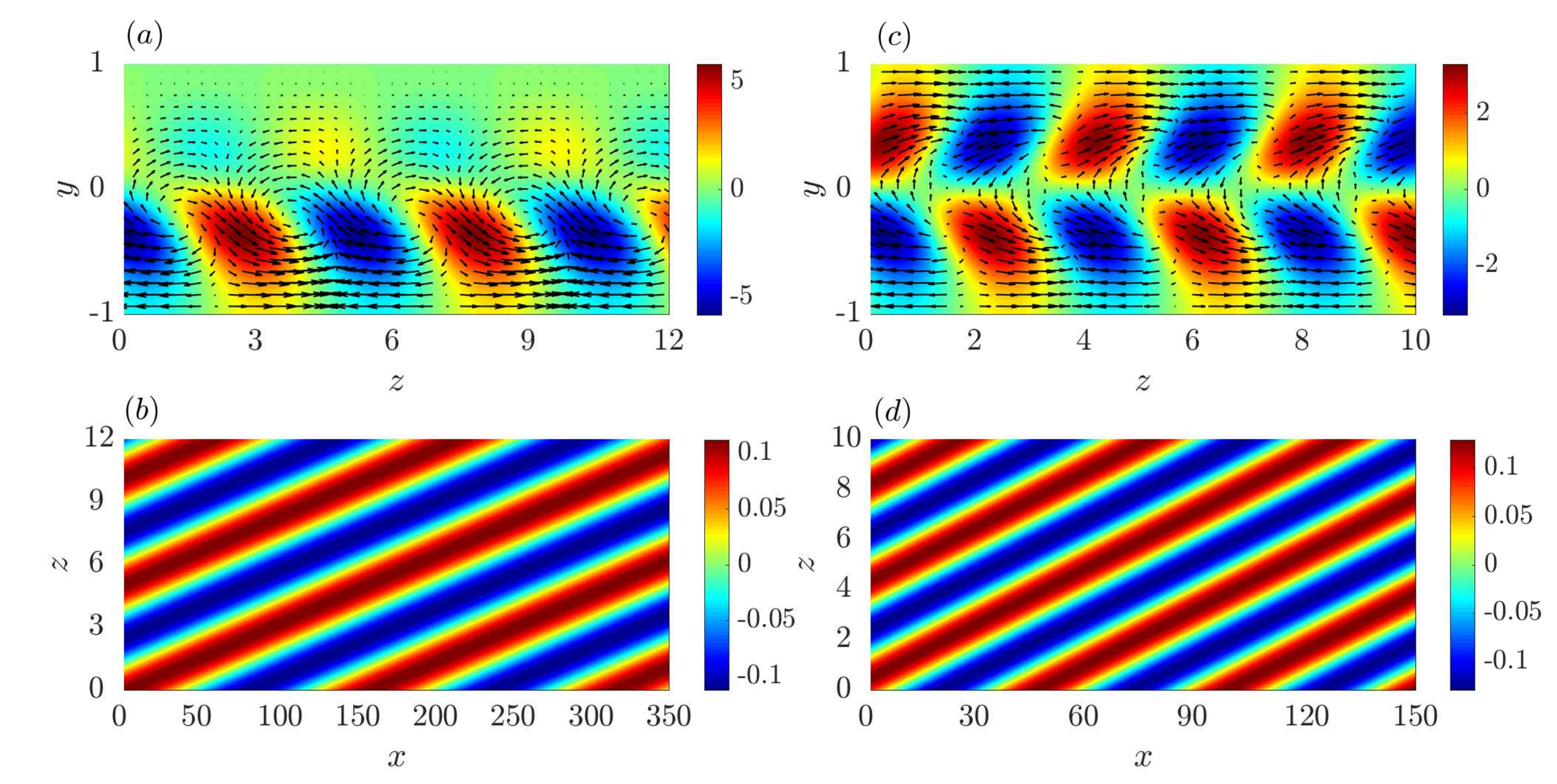}
\caption[wave number]{
    \label{fig:flow_structure_45degree} 
     The flow structure of the leading eigenmode for the one-SH-wall $(a, b)$ and two-SH-wall $(c, d)$ cases with $\theta=\pi/4$ at the respective critical Reynolds number. In $(a, b)$, the bottom wall is slippery, the Reynolds number is $Re=486$ and the critical wavenumbers are $(\alpha, \beta)=(0.04, -1.23)$. In $(c, d)$, the Reynolds number is $Re=429$ and the critical wavenumbers are $(\alpha, \beta)=(0.12, -1.63)$. The color shows the streamwise velocity in the top row and wall-normal velocity in the bottom row, and vectors show the in-plane velocities. The magnitudes of the shown quantities are arbitrary.
} 
\end{figure}

Further, the flow structure of the leading eigenmode at the critical Reynolds number is visualized in figure \ref{fig:flow_structure_0degree} for $\theta=0$ and in figure \ref{fig:flow_structure_45degree} for $\theta=\pi/4$. At $\theta=0$, the flow structure is very similar to the leading eigenmode of the no-slip channel flow, featuring spanwise invariant roll cells filling the whole space between the two walls, except for the region close to the bottom slippery wall. This is reasonable since the top wall is no-slip and therefore the flow structure of the leading eigenmode, at least near the top wall, can be expected to resemble that of the no-slip channel flow. Figure \ref{fig:flow_structure_0degree}(a) shows that the spanwise vorticity is reduced due to the slip at the bottom wall, suggesting that the slip at the bottom wall in fact plays a stabilizing effect with this specific slip setting. This stabilizing effect results in a higher critical Reynolds number compared with the no-slip channel flow (7417 vs. 3848). In the presence of two SH walls, the leading eigenmode is 3-D, exhibiting straight flow structures tilted by an angle about the streamwise direction, see figure \ref{fig:flow_structure_0degree}(d). 
It is noticed from the vector plots in figure \ref{fig:flow_structure_0degree}(b) that the wall-normal velocity component is very small compared to the streamwise component, unlike the no-slip and the one-SH-wall cases where the wall-normal velocity component is comparable to the streamwise one, e.g. see figure \ref{fig:flow_structure_0degree}(a). 

At $\theta=\pi/4$, the leading modes at the respective critical Reynolds numbers in both slip settings are 3-D straight structures (vortices and streaks) tilted about the streamwise direction, see figure \ref{fig:flow_structure_45degree}. In the one-SH-wall case, the flow structures are concentrated in the lower half of the channel, while fluctuations are much weaker close to the upper no-slip wall. This flow structure is similar to that reported in \cite{Pralits2017} where the authors termed it as a wall-vortex mode. It can be seen that the flow structure near the bottom slippery wall closely resembles that in the same flow region in the two-SH-wall case, except for some differences in the wavelengths and flow details. It is noted that the flow features long-streamwise-wavelength (small-$\alpha$) structures in both slip cases and the wavelength in the two-SH-wall case is relatively smaller. Given the similarities in the leading eigenmode in the two cases, it can be expected that the critical Reynolds numbers would be close, just as shown in figure \ref{fig:critical_Re_vs_angle_one_wall}(c). \textcolor{black}{The flow structures in both cases bear some similarities with the unstable modes of the cross-flow instability in 3-D boundary layer flow over swept wings \cite{Mack1984,Saric2003}. For example, the flow exhibits tilted vortices and streaks orientated at small tilt angles about the streamwise direction, i.e. the wave vectors are nearly perpendicular to the streamwise direction ($\arctan{|\frac{\beta}{\alpha}|}\approx 88^\circ$ in the one-SH-wall case (see figure \ref{fig:flow_structure_45degree}b) and $\approx 86^\circ$ in the two-SH-wall case (see figure \ref{fig:flow_structure_45degree}d)).} 

\begin{figure}
\centering
\includegraphics[width=0.99\linewidth]{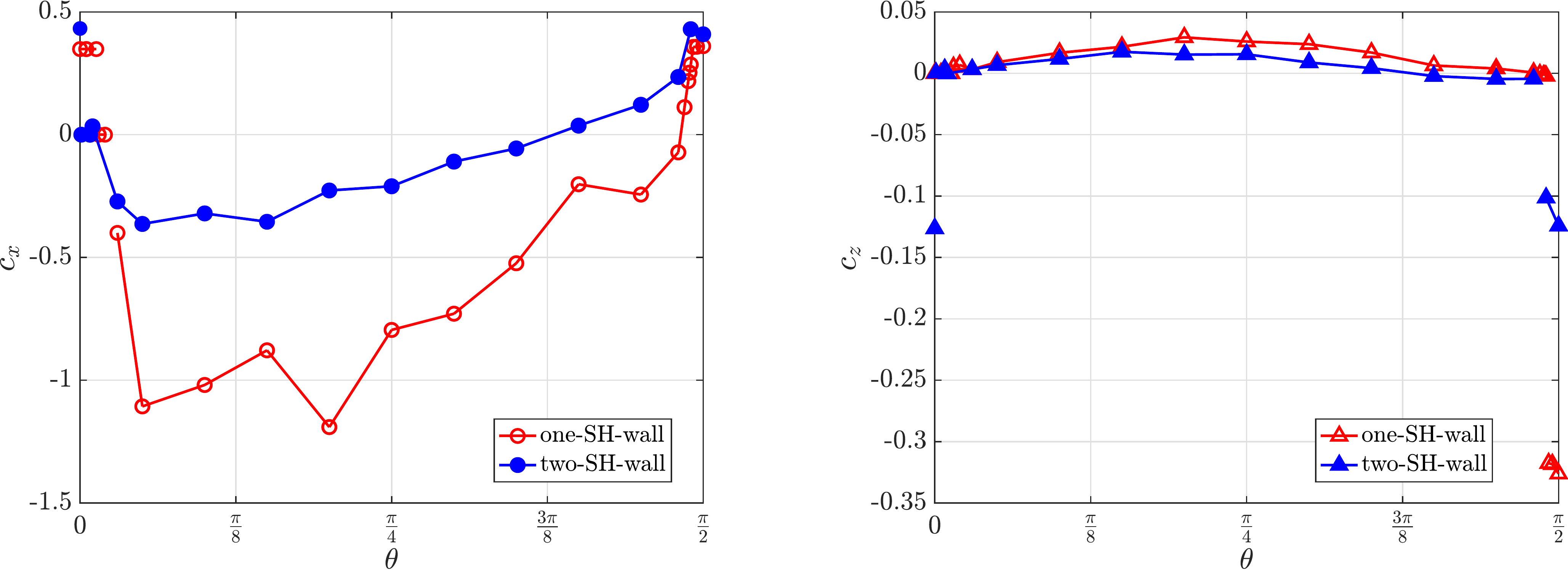}
\caption[wave number]{
    \label{fig:phase_speed} 
     The phase speeds of the leading eigenmode for the $\lambda^\parallel=0.155$ case. The corresponding Reynolds numbers and wavenumbers are shown in figures \ref{fig:critical_Re_vs_angle_one_wall}(c) and \ref{fig:most_unstable_wave_number}. (a) Streamwise phase speed. (b) Spanwise phase speed.
} 
\end{figure}

\textcolor{black}{The streamwise and spanwise phase speeds of the leading eigenmode at the critical Reynolds number with $\lambda^\parallel=0.155$ are shown in figure~\ref{fig:phase_speed}, which are calculated as $c_x=\omega_r/\alpha$ and $c_z=\omega_r/\beta$, respectively, where the frequency $\omega_r$ can be either positive or negative because both forward-propagating and backward-propagating waves are possible. It can be seen that $c_x$ is negative and $c_z$ is positive because $\omega_r$ turns out to be negative away from $\theta=0^\circ$ and $90^\circ$. Because of the small angle $\arctan({\alpha/\beta})$ of the wave structure about the streamwise direction (see figure \ref{fig:flow_structure_45degree}b,d), or in other words, the wave propagates nearly perpendicularly to the streamwise direction, a large backward streamwise propagation speed occurs when a small positive $c_z$ is present.
Close to $\theta=0$ and $\pi/2$, $\omega_r$ is positive and so is $c_x$, and $c_z$ is negative given the opposite signs of $\alpha$ and $\beta$. Because $\alpha$ and $\beta$ are comparable in absolute value, $c_x$ and $c_z$ are also comparable in absolute value, and therefore the wave vector is neither nearly perpendicular nor nearly parallel to the streamwise direction.}

\subsection{Dependence of the critical Reynolds number on $\lambda^\parallel$}
\label{sec:critical_Re_vs_slip_length}

\begin{figure}
\centering
\includegraphics[width=0.9\linewidth]{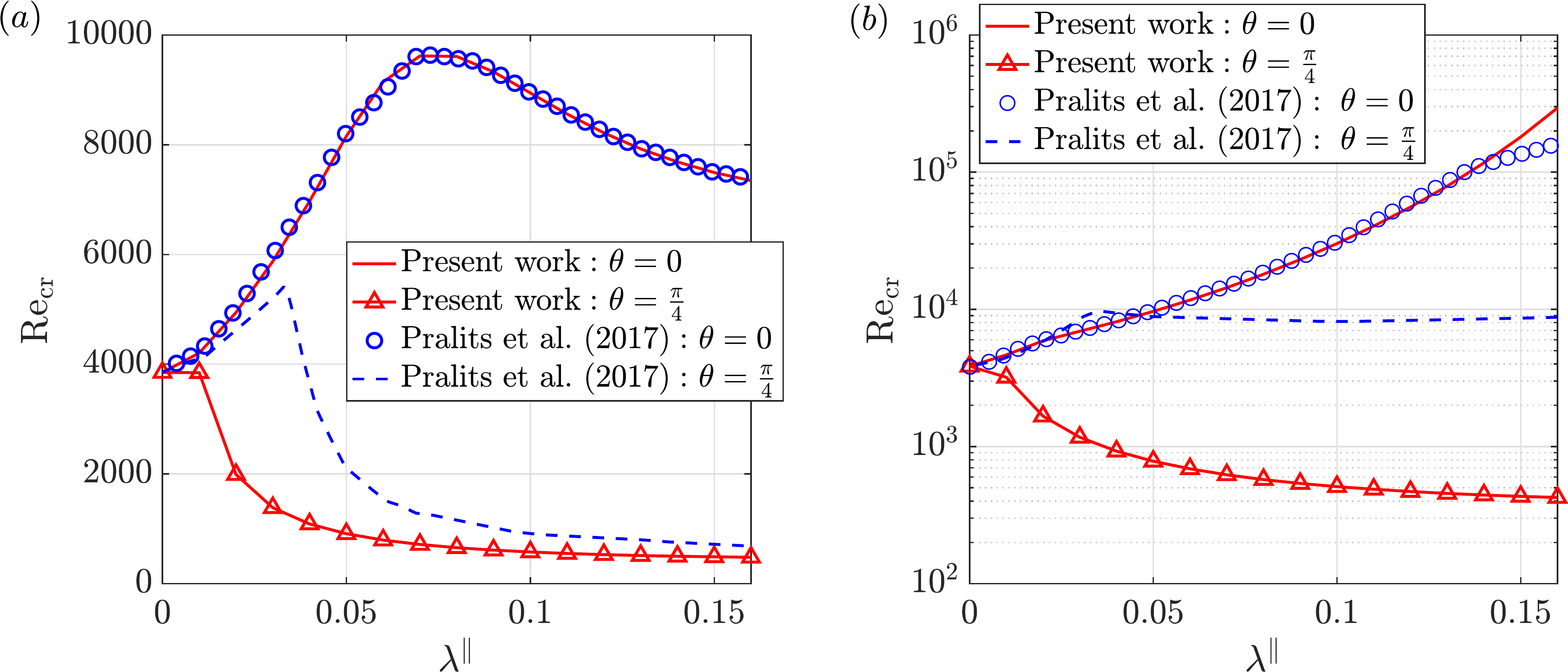}
\caption[wave number]{
    \label{fig:Re_vs_lambda} 
     The critical Reynolds number as a function of $\lambda^\parallel$ with $\theta=0$ and $\theta=\pi/4$. (a) One-SH-wall case, (b) Two-SH-Wall case. The results of \cite{Pralits2017} are also plotted for comparison.
} 
\end{figure}

The slip-length ratio is still fixed at $\lambda^\parallel/\lambda^\bot=2$ in this study as in the section \ref{sec:critical_Re_vs_angle}.
Figure \ref{fig:Re_vs_lambda} shows the critical Reynolds number as a function of the slip length $\lambda^\parallel$ for two special angles $\theta=0$ and $\pi/4$. It can be seen that the $Re_{\mathrm{cr}}$ in our calculation and from \cite{Pralits2017} nearly agree with each other for $\theta=0$, but differ largely for $\theta=\pi/4$ for all $\lambda^\parallel$ values considered.  The difference is more prominent in the two-SH-wall case where our $Re_{\mathrm{cr}}$'s are one-order of magnitude lower than those reported by \cite{Pralits2017}. The trend as $\lambda^\parallel$ increases shows that $Re_{\mathrm{cr}}$ keeps decreasing but gradually levels off at some finite value for $\theta=\pi/4$. For $\theta=0$, at sufficiently large slip length,  $Re_{\mathrm{cr}}$ in the one-SH-wall case monotonically decreases but that in the two-SH-wall case keeps increasing as $\lambda^\parallel$ increases, and the latter becomes orders of magnitude larger than the former as the slip length is large.

\begin{figure}
\centering
\includegraphics[width=0.99\linewidth]{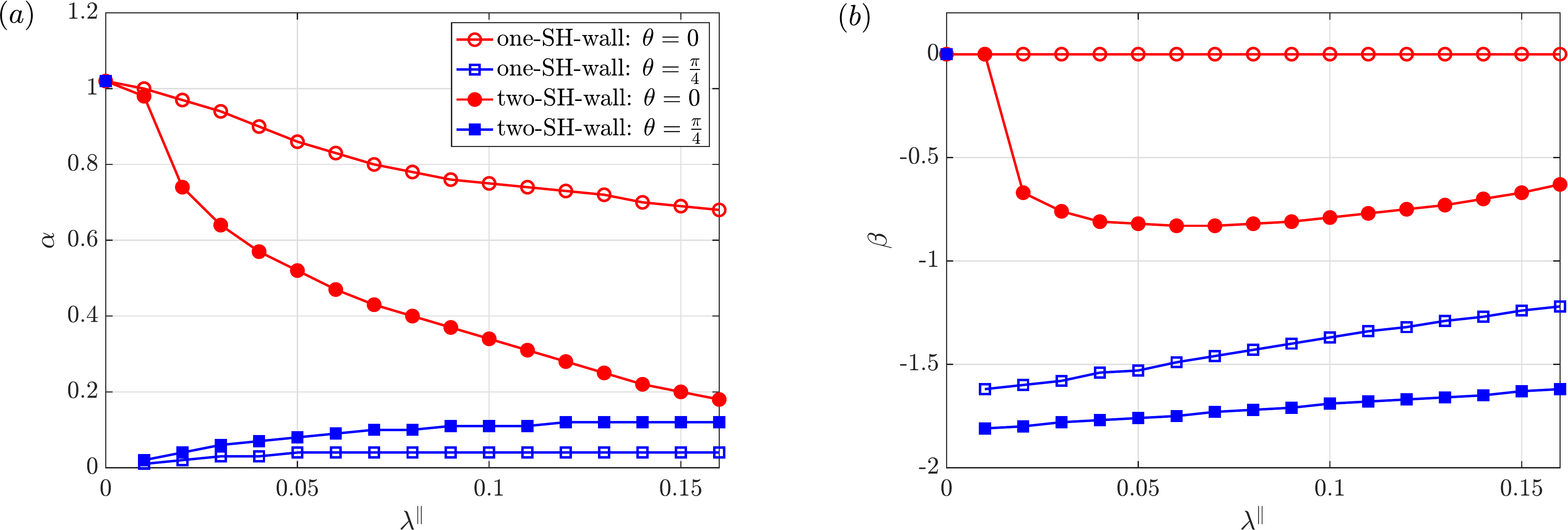}
\caption[wave number]{
    \label{fig:critical_wavenumber_vs_lambda} 
     The critical streamwise (a) and spanwise (b) wavenumbers as functions of $\lambda^\parallel$ with $\theta=0$ and $\theta=\pi/4$.
} 
\end{figure}

\textcolor{black}{The corresponding critical wavenumbers are shown in figure~\ref{fig:critical_wavenumber_vs_lambda}. For $\theta=0$, the critical streamwise wavenumber $\alpha$ keeps decreasing as $\lambda^\parallel$ increases, and $\alpha$ in the two-SH-wall case is smaller and decreases more quickly than the one-SH-wall case. In the one-SH-wall setting, the critical $\beta$ stays as zero for all the $\lambda^\parallel$ values considered, i.e. the flow structure is two-dimensional spanwise invariant as shown in figure~\ref{fig:flow_structure_0degree}(a,b). In contrast, in the two-SH-wall setting, the flow structure is two-dimensional only at very small $\lambda^\parallel$, and the flow becomes three dimensional above $\lambda^\parallel=0.02$. Similar phenomenon was reported in Refs. \cite{Chai2019, Xiong2020}. For $\theta=\pi/4$, instead of a gradual decrease, $\alpha$ drops sharply from 1.02 in the no-slip case to roughly 0.01 and 0.02 at $\lambda^\parallel=0.01$ for the one-SH-wall and two-Sh-wall cases, respectively, and then monotonically increases with $\lambda^\parallel$. Overall, the critical $\alpha$ is much lower than the that with $\theta=0$ for the one-SH-wall case for all $\lambda^\parallel$  considered, whereas the two gradually get close as the $\lambda^\parallel$ increases for the two-SH-wall case. Similar sharp drop also occurs in the critical $\beta$, followed by a gradual increase with $\lambda^\parallel$. Therefore, with $\theta=\pi/4$, the leading eigenmode is three dimensional for both one-SH-wall and two-SH-wall settings even down to $\lambda^\parallel=0.01$. It is noticed that the critical $\beta$ is much larger than the critical $\alpha$ for $\theta=\pi/4$, i.e. the flow structures are of long streamwise wavelengths and much smaller spanwise wavelengths, similar to those shown in figure~\ref{fig:flow_structure_45degree}.}

\begin{figure}
\centering
\includegraphics[width=0.99\linewidth]{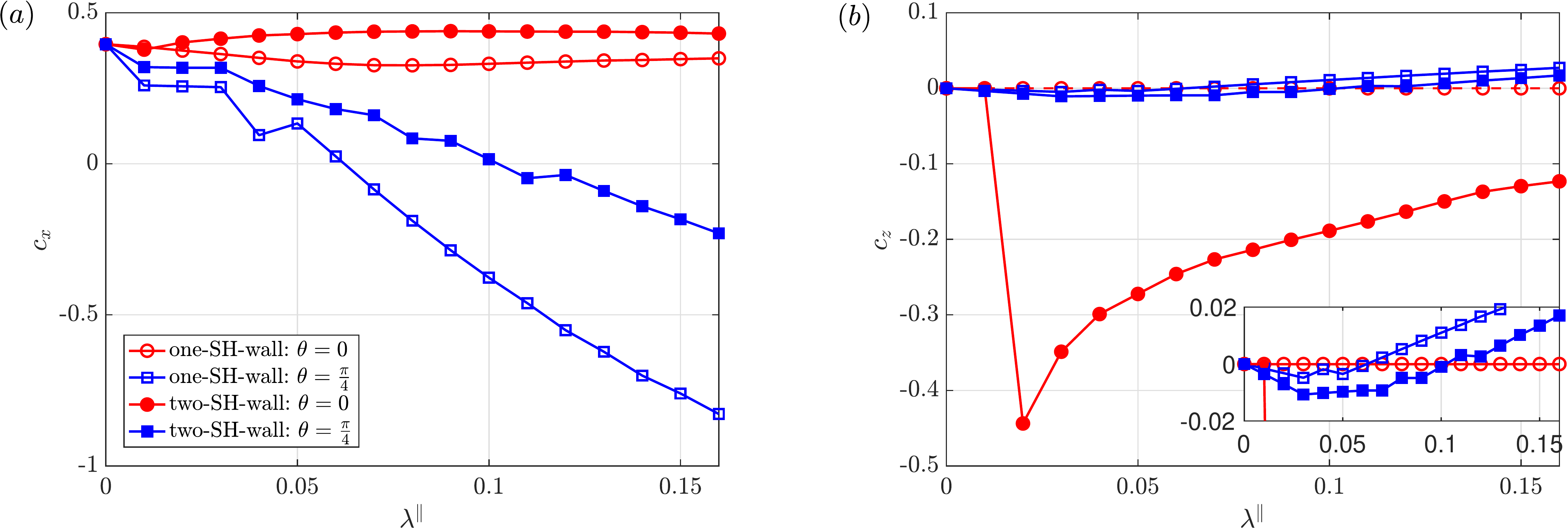}
\caption[phase speed]{
    \label{fig:phase_speed_vs_lambda} 
     The streamwise (a) and spanwise (b) phase speeds of the most unstable eigenmodes as functions of $\lambda^\parallel$ with $\theta=0$ and $\theta=\pi/4$.
} 
\end{figure}

\textcolor{black}{The corresponding critical phase speed of the leading eigenmode as a function of $\lambda^\parallel$ is also shown in figure~\ref{fig:phase_speed_vs_lambda}. For $\theta=0$, the streamwise phase speed $c_x$ stays positive and nearly constant for both one-SH-wall and two-SH-wall cases. The spanwise phase speed $c_z$ stays zero because $\beta$ stays zero in the one-SH-wall case, whereas in the two-SH-wall case, it stays zero at small $\lambda^\parallel$ but then sharply drops to a negative value and remains negative as $\lambda^\parallel$ increases further. This trend is determined by the trend in $\beta$ as shown in figure~\ref{fig:critical_wavenumber_vs_lambda}(b).
For $\theta=\pi/4$,
$c_x$ gradually decreases and becomes negative as $\lambda^\parallel$ is sufficiently large with $\theta=\pi/4$ for both slip settings. This indicates that the wave becomes back-propagating against the base flow in the streamwise direction at sufficiently large slip lengths for $\theta=\pi/4$. As $\alpha$ stays positive, this sign switch is due to the sign switch of the frequency $\omega_r$. Consistent with the trend in $\omega_r$, $c_z$ is firstly negative and then becomes positive as $\lambda^\parallel$ increases (see the inset in figure~\ref{fig:phase_speed_vs_lambda}b), given that the signs of $\alpha$ and $\beta$ are opposite for both slip settings, as seen in figure~\ref{fig:critical_wavenumber_vs_lambda}. Similarly, this indicates that the wave is firstly backward-propagating and then turns forward-propagating in the spanwise direction as $\lambda^\parallel$ increases.}

\subsection{The critical Reynolds number with larger $\lambda^\parallel/\lambda^\bot$}

The slip length ratio $\lambda^\parallel/\lambda^\bot=2$ Ref. \cite{Pralits2017} used is based on the theoretical work of \cite{Lauga2003, Asmolov2012, Philip1972} for one-dimensional texture modeled by periodic alternating no-slip and shear-free regions on the wall. This ratio can be considered as a measure of the anisotropy in the slip length. In this section, we want to investigate the influence of this ratio on the stability of the flow, particularly of larger ratios. In fact, some studies suggested that this ratio can be significantly increased if the liquid is allowed to partially penetrate into the grooves that contain gas pockets. For example, Ref.  \cite{Ng2009} showed that the transverse slip length $\lambda^\bot$ is more sensitive to the penetration (decreases more quickly with increasing penetration) than the longitudinal slip length $\lambda^\parallel$, and reported a ratio of up to 4. 

\begin{figure}
\centering
\includegraphics[width=0.99\linewidth]{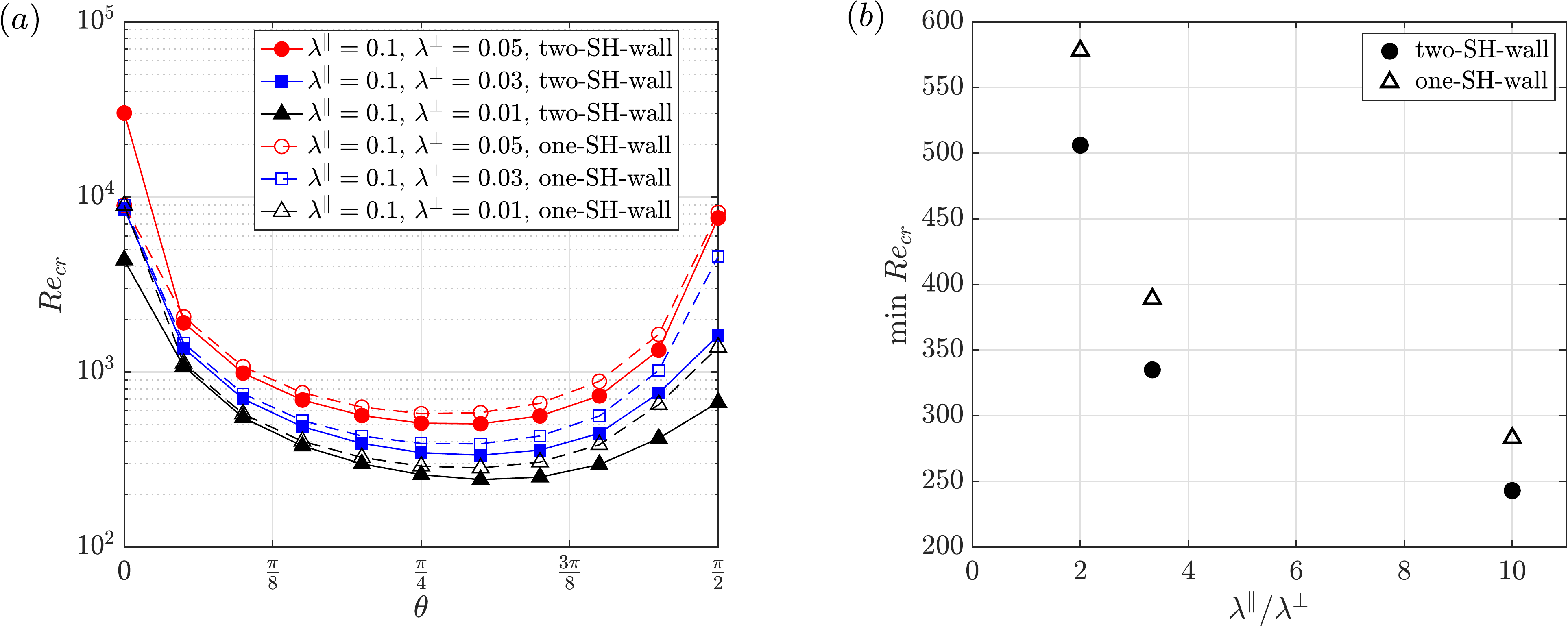}
\caption[large ratio]{
    \label{fig:critical_Re_large_ratio}
     (a) The critical Reynolds number as a function of the tilt angle $\theta$ for the one-SH-wall (solid lines) and two-SH-wall (dashed lines) with $\lambda^\parallel=0.1$ and $\lambda^\parallel/\lambda^\bot=2$, $10/3$ and 10.
(b) The minimum critical Reynolds number over $\theta$ as a function of the slip length ratio $\lambda^\parallel/\lambda^\bot$. } 
\end{figure}

Figure \ref{fig:critical_Re_large_ratio} shows the critical Reynolds number as a function of the tilt angle $\theta$ for a few larger values of $\lambda^{\parallel}/\lambda^{\bot}$ \textcolor{black}{up to 10}. Fixing $\lambda^\parallel=0.1$, we consider $\lambda^\bot=0.05$, 0.3 and 0.01. It can be seen that the critical Reynolds number decreases for all $\theta$ as the ratio increases. The critical Reynolds number can be reduced to about 243 for the two-SH-wall case when the ratio is increased to 10. Ref. \cite{Pralits2017} analyzed the applicability of the slip boundary condition with respect to the value of the slip length. Their analysis suggested that, 
the upper limit of the non-dimensional slip length with which the boundary condition \eqref{equ:BC2} still applies is roughly 0.1. 
Overall, the two-SH-wall setting nearly always gives lower $Re_{cr}$ than that given by the one-SH-wall setting for all  slip-length ratios considered, except for very small $\theta$ values. 

	\begin{table}[h!]
	\begin{center}
		\begin{tabular}{|c|c|c|c|}
			\hline
			\multicolumn{2}{|c|}{parameters} & $\theta_{1}$ & $\theta_{2}$\\
			\hline
			
			$\lambda^{\parallel}=0.05,\lambda^{\perp}=0.025$ & two-SH-wall & $0.26\pi$ & $0.27\pi$\\
			\hline
			
			$\lambda^{\parallel}=0.07,\lambda^{\perp}=0.035$ & one-SH-wall & $0.25\pi$ & $0.26\pi$\\
			\hline
					
			\multirow{2}{*}{$\lambda^{\parallel}=0.155,\lambda^{\perp}=0.0775$} & one-SH-wall & $0.26\pi$ & $0.27\pi$\\
			\cline{2-4}
			& two-SH-wall & $0.26\pi$ & $0.28\pi$\\
			\hline

			\multirow{2}{*}{$\lambda^{\parallel}=0.1,\lambda^{\perp}=0.05$} & one-SH-wall & $0.26\pi$ & $0.27\pi$\\
            \cline{2-4}
            & two-SH-wall & $0.26\pi$ & $0.28\pi$\\
            \hline			

			\multirow{2}{*}{$\lambda^{\parallel}=0.1,\lambda^{\perp}=0.03$} & one-SH-wall & $0.26\pi$ & $0.27\pi$\\
            \cline{2-4}
            & two-SH-wall & $0.26\pi$ & $0.29\pi$\\
            \hline	
			    
			\multirow{2}{*}{$\lambda^{\parallel}=0.1,\lambda^{\perp}=0.01$} & one-SH-wall & $0.26\pi$ & $0.28\pi$\\
            \cline{2-4}
            & two-SH-wall & $0.27\pi$ & $0.3\pi$\\
            \hline	


			\end{tabular}
			\caption{\label{tab:correlation_with_crossflow}Comparison of the tilt angle corresponding to the maximum spanwise velocity component in the base flow, $\theta_1$, and the tilt angle corresponding to the lowest critical Reynolds number, $\theta_2$, at different parameter settings. }
		\end{center}

	\end{table}

We also compare with the limiting case of pure spanwise slip on both walls considered by \cite{Chai2019}, corresponding to the two-SH-wall case with $\theta=\pi/2$, $\lambda^\parallel=\lambda_z$ being finite and $\lambda^\bot=0$ ($\lambda^\parallel/\lambda^\bot\to \infty$) in the present slip setting. For $\lambda_z=0.1$, the critical Reynolds number given by \cite{Chai2019} is 489. \textcolor{black}{However, an infinite slip-length ratio is certainly unrealistic and the ratio necessarily remains finite in experiments. For a finite slip length ratio, our results at $\theta=90^\circ$ show higher critical Reynolds numbers compared to the limiting case of Ref.~\cite{Chai2019}. Nevertheless, a proper tilt angle can greatly reduce the critical Reynolds number.} As shown in figure \ref{fig:critical_Re_large_ratio}(b), the lowest critical Reynolds numbers are $Re_{\mathrm{cr}}\approx 389$ for the one-SH-wall case with a ratio of $10/3$ and $Re_{\mathrm{cr}}\approx 283$ with a ratio of 10 at a tilt angle of $\theta\approx 54^\circ$. The numbers for the two-SH-wall case are $Re_{\mathrm{cr}}\approx 335$ and 243 for the two ratios at $\theta\approx 54^\circ$, respectively. Therefore, compared to the pure spanwise slip case, lower critical Reynolds numbers can be achieved by using proper tilt angles of the microgrooves with finite slip-length ratios.

	
	\textcolor{black}{Collecting all the data, here we show the correlation between the instability and the cross-flow $W$ by comparing the tilt angle that minimizes the critical Reynolds number and that maximizes the magnitude of $W$, see table \ref{tab:correlation_with_crossflow}. It can be seen that for all the cases with $\lambda^{\parallel}/\lambda^{\bot}=2$, the two angles are very close to each other. For larger ratios with $\lambda^\parallel=0.1$, the deviation between the two angles seems to increase as the ratio increases, especially for the two-SH-wall setting. 
Overall, the two angles are rather close to each other, indicating a strong correlation between the instability and the magnitude of the cross-flow component in the base flow caused by the tilt angle of the microgrooves. This correlation can also be seen by comparing figure \ref{fig:baseflow}(a) and figure \ref{fig:critical_Re_vs_angle_one_wall}(c).}

\subsection{With different tilt angles at the two walls}

In Ref. \cite{Pralits2017} and in our previous sections, the microgrooves on the top and bottom walls were assumed to be parallel to each other, i.e. the tilt angle $\theta$ is identical on both walls. In this section, we investigate the case when $\theta$ differs at the two walls. We did not derive the analytical basic flow for this non-parallel case, instead, we numerically solve for the basic flow from the governing equations. The following growth rate calculations are performed using the DNS approach as described in section \ref{sec:methods} and Appendix \ref{sec:NS_solver}.

\begin{figure}
\centering
\includegraphics[width=0.9\linewidth]{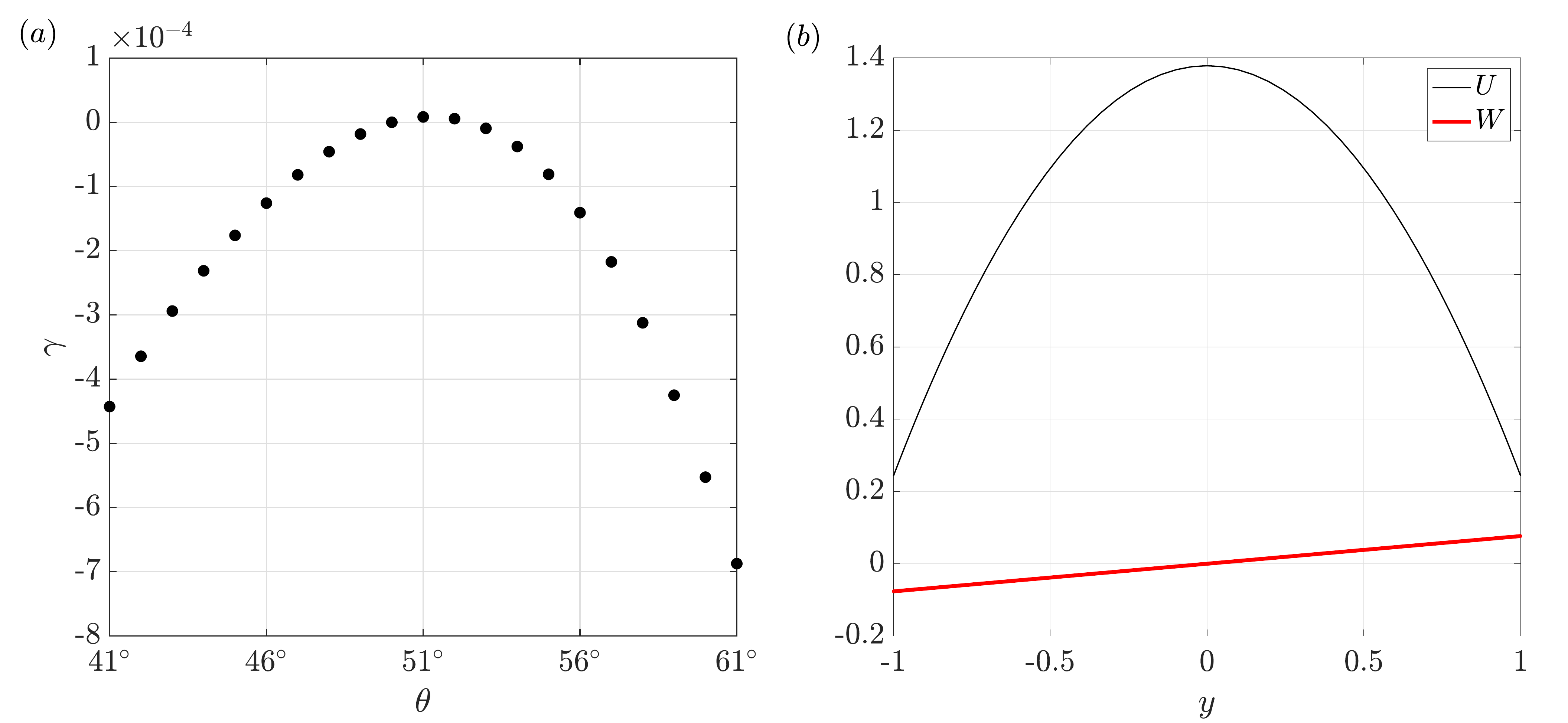}
\caption[wave number]{
    \label{fig:non_parallel} 
     (a) The growth rate $\gamma$ of the most unstable mode for the two-SH-wall case with different tilt angle $\theta$ on the two walls. Slip lengths are $\lambda^\parallel=0.155$ and $\lambda^\bot=\lambda^\parallel/2$. The angle is fixed at $\theta=51^{\circ}$ on the top wall and  is varied on the bottom wall. The Reynolds numbers is fixed at  $Re=419$. (b) The base flow profiles $U$ and $W$ in case of $\theta=51^\circ$ on top wall and $\theta=-51^\circ$ on the bottom wall. Slip lengths are the same as in panel (a).
} 
\end{figure}

For this study, we choose the two-SH-wall setting that gives the lowest critical Reynolds number $Re_{\mathrm{cr}}=419$ with $\theta=51^\circ$ on both walls. We fix the tilt angle $\theta=51^\circ$ on the top wall while change $\theta$ at the bottom wall. Instead of searching for the critical Reynolds number directly, we fix the Reynolds number to be $Re=419$ and calculate the largest growth rate by scanning the wavenumber plane. In this approach, a larger maximum growth rate would indicate a lower critical Reynolds number.

Figure \ref{fig:non_parallel}(a) shows the maximum growth rate for a few tilt angles at the bottom wall. The data show that the growth rate peaks at $\theta=51^\circ$, which is equal to the fixed tilt angle at the top wall. The flow becomes linearly stable when the angles on the two walls deviate from each other, and the trend shows a monotonic decrease as the difference increases, indicating that the lowest critical Reynolds number is realized in the equal tilt angle setting, i.e. with parallel microgrooves on the two walls. We also tested an extreme case where $\theta=-51^\circ$ on the bottom wall, for which the base flow profiles are shown in figure \ref{fig:non_parallel}(b). We obtained a growth rate of $-0.0036$ for the least stable mode, which is much lower than those shown in figure \ref{fig:non_parallel}(a), i.e. the flow is much more stable.

\section{Conclusions}
The destabilizing effect of anisotropic slip of microgrooves, modeled by a tensorial slip boundary condition, on channel flow is studied in this paper. Our results agree with Ref. \cite{Pralits2017} that a proper tilt angle in the microgrooves about the streamwise direction can significantly reduce the critical Reynolds number for the onset of linear instability. \textcolor{black}{With a proper tilt angle, the destabilizing effect of the slip is already noticeable at a small slip length of $\lambda^\parallel=0.01$, see figure \ref{fig:Re_vs_lambda}b. The instability seems to be related to the cross flow caused by the tilt angle because the reduction in the critical Reynolds number is strongly correlated with the magnitude of the the cross-flow component of the base flow. The instability bears some similarities with the cross-flow instability of 3-D boundary layer flow over swept wings \cite{Mack1984,Saric2003}. However, the mechanism here may be different from the latter in that the cross-flow component of the base flow here is not inflectional (either constant or linear) but is inflectional in swept flows, as also pointed out by Ref. \cite{Jouin2022}. Besides, 3-D leading instability can occur even if the cross-flow component of the base flow vanishes (see figure \ref{fig:unstable_region}b, figure \ref{fig:flow_structure_0degree}c,d and Refs. \cite{Chai2019,Xiong2020}). The destabilizing mechanism of the anisotropic slip remains open.}

However, overall our critical Reynolds numbers are much lower than those reported by Ref. \cite{Pralits2017}. In contrast to the conclusion of \cite{Pralits2017}, our results show that the two-SH-wall setting nearly always results in a lower critical Reynolds number compared to the one-SH-wall setting (see figure \ref{fig:critical_Re_vs_angle} and \ref{fig:critical_Re_vs_angle_one_wall}).
An exception is at very small $\theta$ where the former may result in higher critical Reynolds numbers for some specific slip length settings, as in the $\lambda^\parallel=0.155$ and $\frac{\lambda^\parallel}{\lambda^\bot}=2$ case shown in figure \ref{fig:critical_Re_vs_angle_one_wall}(c).  The critical Reynolds number can be further reduced if the anisotropy in the slip length is increased, see figure \ref{fig:critical_Re_large_ratio}.
These results are cross-validated by using three different formulations for the eigenvalue calculation.

The results suggest that, if instability is preferred, tilt angles close to $0^\circ$ and $90^\circ$ should generally be avoided. \textcolor{black}{The tilt angle that maximizes the cross-flow component of the base flow nearly gives the lowest critical Reynolds number and therefore is recommended.} The microgrooves should be parallel on top and bottom walls for realizing the lowest critical Reynolds number, because the flow becomes more stable/less unstable when the difference in the angles at the two walls increases, see figure \ref{fig:non_parallel}.
Besides resulting in lower critical Reynolds numbers, the two-SH-wall setting can cause instability at larger streamwise and spanwise wavenumbers (see figure \ref{fig:most_unstable_wave_number}), i.e. shorter wavelengths, compared with the one-SH-wall setting, which may be preferable in applications that require enhancing mixing. Besides, shorter wavelengths also pose less restriction on the channel size in experiments if instability is to be induced.


\section{Acknowledgements}
The authors acknowledge financial support from the National Natural Science Foundation of China under the grant number 91852105 and from Tianjin University under the grant number 2018XRX-0027.
 

\bibliographystyle{unsrt}
\bibliography{references}

\appendix
\section{The Navier-Stokes solver}\label{sec:NS_solver}
The nondimensional linearized Navier-Stokes equations were given in the main text as \eqref{equ:LNS}. However, for the ease of presentation here, we repeat them as following,

\begin{equation}
	\label{the linear Navier-Stokes equation}
	\frac{\partial \boldsymbol{u}}{\partial t}+\boldsymbol{U}\cdot \nabla\boldsymbol{u}+\boldsymbol{u}\cdot \nabla\boldsymbol{U}=-\nabla p+\frac{1}{Re}\nabla^{2}\boldsymbol{u}\,,    
\end{equation}
\begin{equation}
	\label{divergence-free condition}
	\nabla\cdot\boldsymbol{u} = 0\,,
\end{equation}
where $\boldsymbol{U}$ denotes the steady base flow, $\boldsymbol{u}$ and $p$ denote perturbative velocity and disturbance, respectively. 

An semi-implicit second-order Adams-Bashforth/backward differentiation
scheme \cite{Hugues1998} is used for time integration. After the temporal discretization, Eq.\eqref{the linear Navier-Stokes equation} is rearranged into
\begin{equation}
\label{temporal discretization}
\frac{3\boldsymbol{u}^{n+1}-4\boldsymbol{u}^{n}+\boldsymbol{u}^{n-1}}{2\Delta t}+2\boldsymbol{N} (\boldsymbol{u}^{n})-\boldsymbol{N} (\boldsymbol{u}^{n-1})=
-\nabla p ^{n+1}+\frac{1}{Re}\nabla ^{2}\boldsymbol{u}^{n+1}, 
\end{equation}
where $n-1,n$ and $n+1$ are the indices for the previous, current and next time steps, $ \Delta t$ is the time step size and
\begin{equation}
\label{definition of nonlinear term N}
\boldsymbol{N}(\boldsymbol{u} ) :=\boldsymbol{U}\cdot \nabla \boldsymbol{u}+\boldsymbol{u}\cdot \nabla\boldsymbol{U}
\end{equation}
is the advection term (the nonlinear convection term $\boldsymbol{u}\cdot\nabla\boldsymbol{u}$ can be included here for nonlinear simulations). We set $\boldsymbol{u}^{-1}=\boldsymbol{u}^{0}$, where $\boldsymbol{u}^{0}$ is the initial condition. Equation \eqref{temporal discretization} can be rewritten as
\begin{equation}
\label{Operator form of temporal discretization}
L\boldsymbol{u}^{n+1}=Re\nabla p^{n+1}-\frac{2Re}{\Delta t}\boldsymbol{u}^{n}+\frac{Re}{2\Delta t} \boldsymbol{u}^{n-1}+2Re\boldsymbol{N} (\boldsymbol{u}^{n})-Re\boldsymbol{N} (\boldsymbol{u}^{n-1}),
\end{equation} 
where
\begin{equation}
\label{The operator acting on the velocity of the (n+1)th step}
L :=\nabla^{2}-\frac{3Re}{2\Delta t}\,.
\end{equation}

Suppose that the velocity field at the $n^{th}$ time step is divergence-free, i.e., $\nabla\cdot \boldsymbol{u}^{n}=0$,  taking the divergence of Eqs. \eqref{temporal discretization}, we obtain
\begin{multline}
\label{the divergence of temporal discretization}
\frac{3\nabla\cdot \boldsymbol{u}^{n+1}-4\nabla\cdot \boldsymbol{u}^{n}+\nabla\cdot \boldsymbol{u}^{n-1}}{2\Delta t}+2\nabla\cdot \boldsymbol{N} (\boldsymbol{u}^{n})-\nabla\cdot \boldsymbol{N} (\boldsymbol{u}^{n-1})=\\
-\nabla^{2}  p^{n+1}+\frac{1}{Re}\nabla^{2} (\nabla\cdot\boldsymbol{u}^{n+1}).
\end{multline}
Let $\boldsymbol{\nu}=\nabla\cdot \boldsymbol{u}^{n+1}$, Eqs.~\eqref{the divergence of temporal discretization} can be rewritten as
\begin{equation}
\label{Another form of the divergence of temporal discretization}
\frac{3}{2\Delta t}\boldsymbol{\nu}-\frac{1}{Re}\nabla^{2}\boldsymbol{\nu}=-\nabla^{2} p^{n+1}-2\nabla\cdot \boldsymbol{N} (\boldsymbol{u}^{n})+\nabla\cdot \boldsymbol{N} (\boldsymbol{u}^{n-1})+\frac{2}{\Delta t} \nabla\cdot \boldsymbol{u}^{n}-\frac{1}{2\Delta t}\nabla\cdot \boldsymbol{u}^{n-1}\,.
\end{equation}
If let $p^{n+1}$ satisfy:
\begin{equation}
\label{equation for pressure}
\nabla^{2}  p^{n+1}=-2\nabla\cdot \boldsymbol{N} (\boldsymbol{u}^{n})+\nabla\cdot \boldsymbol{N} (\boldsymbol{u}^{n-1})+\frac{2}{\Delta t} \nabla\cdot \boldsymbol{u}^{n}-\frac{1}{2\Delta t}\nabla\cdot \boldsymbol{u}^{n-1}\,,
\end{equation}
then
\begin{equation}
\label{useless equation 1}
\frac{3}{2\Delta t}\boldsymbol{\nu}-\frac{1}{Re}\nabla^{2}\boldsymbol{\nu}=0\,.
\end{equation}
According to the maximum principle \cite{Phillips1991}, if we impose $\boldsymbol{\nu}\equiv 0$ on the boundary $y=\pm1$, then $\boldsymbol{\nu}\equiv 0$ within the whole flow domain.
Therefore, to make the solution $\boldsymbol u^{n+1}$ satisfies $\nabla\cdot \boldsymbol{u}^{n+1}=0$, it is only necessary to solve Eqs.~\eqref{equation for pressure},
where the boundary condition for $p^{n+1}$ makes $\nabla\cdot \boldsymbol{u}^{n+1}=0$ hold on the wall boundary, besides the slip boundary condition \eqref{equ:BC2}. In the following, we show the influence matrix technique for satisfying the required boundary conditions.

Suppose that we have the solutions at the time steps $n$ and $n-1$, we follow the following procedure to obtain the solutions at time step $n+1$.

1.Let $\overline{p}^{n+1}$ be the solutions of Eq.\eqref{equation for pressure} with the homogeneous Neumann boundary conditions, i.e.
\begin{equation}
\label{predicted pressure}
\left\{\begin{matrix} 
\nabla^{2}  \overline{p}^{n+1} & = & -2\nabla\cdot \boldsymbol{N} (\boldsymbol{u}^{n})+\nabla\cdot \boldsymbol{N} (\boldsymbol{u}^{n-1})+\dfrac{2}{\Delta t} \nabla\cdot \boldsymbol{u}^{n}-\dfrac{1}{2\Delta t}\nabla\cdot \boldsymbol{u}^{n-1} \\  

\dfrac{\partial \overline{p}^{n+1}}{\partial y}  & = & \hspace{-15em} 0 ,\qquad \text{at} \quad y=\pm 1.
\end{matrix}\right. 
\end{equation}
Let $\overline{\boldsymbol{u}} ^{n+1}$ be the solutions of Eqs.~\eqref{Operator form of temporal discretization} with the homogeneous Dirichlet boundary conditions. Using $\overline{p}^{n+1}$, a prediction $\overline{\boldsymbol{u}} ^{n+1}$ can be obtained by solving
\begin{equation}
\label{predicted velocity}
\left\{\begin{matrix} 
L\overline{\boldsymbol{u}} ^{n+1} & = & Re\nabla \overline{p}^{n+1}-\dfrac{2Re}{\Delta t}\boldsymbol{u}^{n}+\dfrac{Re}{2\Delta t}\boldsymbol{u}^{n-1}+2Re\boldsymbol{N} (\boldsymbol{u}^{n})-Re\boldsymbol{N} (\boldsymbol{u}^{n-1}) \\  

\overline{\boldsymbol{u}} ^{n+1} & = & \hspace{-15em} 0 ,\qquad \text{at} \quad y=\pm 1.
\end{matrix}\right. 
\end{equation}
Note that $\overline{\boldsymbol{u}}^{n+1}$ satisfies neither boundary condition \eqref{equ:BC2} nor the divergence free condition at the channel walls.

2.
We define several basis functions, which can adjust the velocity on the boundary without affecting the solution in the bulk. Let $\boldsymbol{u}_{i}^{\dagger}, i=1,2$, satisfy the following equations and boundary conditions
\begin{equation}
\label{basis functions of the streamwise velocity components}
\begin{cases}
Lu_{1}^{\dagger}=0\\
u_{1}^{\dagger}(x,+1,z)=0\\
u_{1}^{\dagger}(x,-1,z)=1\\
v_{1}^{\dagger}(x,y,z)=0\\
w_{1}^{\dagger}(x,y,z)=0
\end{cases}
\hspace{5em}
\begin{cases}
Lu_{2}^{\dagger}=0\\
u_{2}^{\dagger}(x,+1,z)=1\\
u_{2}^{\dagger}(x,-1,z)=0\\
v_{2}^{\dagger}(x,y,z)=0\\
w_{2}^{\dagger}(x,y,z)=0
\end{cases}.
\end{equation}
Similarly,  $\boldsymbol{u}_{i}^{\dagger}, i=3,4,5,6$ can be constructed to satisfy
\begin{equation}
\label{basis functions of the wall-normal velocity components}
\begin{cases}
u_{3}^{\dagger}(x,y,z)=0\\
Lv_{3}^{\dagger}=0\\
v_{3}^{\dagger}(x,+1,z)=0\\
v_{3}^{\dagger}(x,-1,z)=1\\
w_{3}^{\dagger}(x,y,z)=0
\end{cases}
\hspace{5em}
\begin{cases}
u_{4}^{\dagger}(x,y,z)=0\\
Lv_{4}^{\dagger}=0\\
v_{4}^{\dagger}(x,+1,z)=1\\
v_{4}^{\dagger}(x,-1,z)=0\\
w_{4}^{\dagger}(x,y,z)=0
\end{cases}.
\end{equation}

\begin{equation}
\label{basis functions of the spanwise velocity components}
\begin{cases}
u_{5}^{\dagger}(x,y,z)=0\\
v_{5}^{\dagger}(x,y,z)=0\\
Lw_{5}^{\dagger}=0\\
w_{5}^{\dagger}(x,+1,z)=0\\
w_{5}^{\dagger}(x,-1,z)=1\\
\end{cases}
\hspace{5em}
\begin{cases}
u_{6}^{\dagger}(x,y,z)=0\\
v_{6}^{\dagger}(x,y,z)=0\\
Lw_{6}^{\dagger}=0\\
w_{6}^{\dagger}(x,+1,z)=1\\
w_{6}^{\dagger}(x,-1,z)=0\\
\end{cases}.
\end{equation}
Note that the boundary conditions for $p^{n+1}$ must make $\nabla\cdot \boldsymbol{u}^{n+1}=0$ also hold on the boundary. There are eight boundary conditions on the two walls in total, and two more basis functions associated with the pressure are needed. Following \cite{Willis2017}, we construct $\boldsymbol{u}_{i}^{\dagger}, i=7,8$ as the following:
\begin{equation}
\label{basis functions}
\begin{cases}
\begin{cases}
\nabla^{2}{p}^{\dagger}=0\\
\dfrac{\partial {p}^{\dagger}}{\partial y}(x,+1,z)=0\\
\dfrac{\partial {p}^{\dagger}}{\partial y}(x,-1,z)=1\\
\end{cases}\\
\boldsymbol{u} _{7}^{\dagger}(x,y,z)=-\nabla{p}^{\dagger}
\end{cases}
\hspace{5em}
\begin{cases}
\begin{cases}
\nabla^{2}{p}^{\dagger}=0\\
\dfrac{\partial {p}^{\dagger}}{\partial y}(x,+1,z)=1\\
\dfrac{\partial {p}^{\dagger}}{\partial y}(x,-1,z)=0\\
\end{cases}\\
\boldsymbol{u} _{8}^{\dagger}(x,y,z)=-\nabla{p}^{\dagger}
\end{cases}\quad.
\end{equation}
which can be used to adjust the pressure gradient at the boundary without affecting the right hand side of the pressure Poisson equation in the bulk.

The desired solution $\boldsymbol{u}^{n+1}$ can be constructed as 
\begin{equation}
\label{linear combination of solution}
\boldsymbol{u}^{n+1}=\overline{\boldsymbol u}^{n+1}+\sum_{i=1}^{8}a_{i} \boldsymbol{u} _{i}^{\dagger}\,,
\end{equation}
where $a_i$'s are coefficients to be determined using the eight (four on each wall) boundary conditions for $\boldsymbol u^{n+1}$, i.e. Eqs.~\eqref{equ:BC2}, impermeability condition $v=0$ and the divergence free condition $\nabla\cdot\boldsymbol u^{n+1}=0$ at the boundary $y=\pm 1$.

Specifically, the equations for the coefficients $a_i$'s read
\begin{equation}
\label{solution of coefficients}
\begin{cases}
\begin{bmatrix}
\sum_{i=1}^{8}a_{i} u_{i}(x,-1,z)\\\sum_{i=1}^{8}a_{i} w_{i}(x,-1,z)
\end{bmatrix}
=\boldsymbol{\Lambda}\dfrac{\partial}{\partial y}
\begin{bmatrix}
(\overline u^{n+1}+\sum_{i=1}^{8}a_{i} u_{i}^{\dagger})(x,-1,z)\\(\overline w^{n+1}+\sum_{i=1}^{8}a_{i} w_{i}^{\dagger})(x,-1,z)
\end{bmatrix}
\vspace{0.5em}
\\
\begin{bmatrix}
\sum_{i=1}^{8}a_{i} u_{i}(x,+1,z)\\\sum_{i=1}^{8}a_{i} w_{i}(x,+1,z)
\end{bmatrix}
=-\boldsymbol{\Lambda}\dfrac{\partial}{\partial y}
\begin{bmatrix}
(\overline u^{n+1}+\sum_{i=1}^{8}a_{i} u_{i}^{\dagger})(x,+1,z)\\(\overline w^{n+1}+\sum_{i=1}^{8}a_{i} w_{i}^{\dagger})(x,+1,z)
\end{bmatrix}
\vspace{0.5em}
\\
\sum_{i=1}^{8}a_{i} v_{i}(x,+1,z)=0 \vspace{0.5em}
\\
\sum_{i=1}^{8}a_{i} v_{i}(x,-1,z)=0 \vspace{0.5em}
\\
\quad\nabla\cdot(\overline{\boldsymbol u}^{n+1}+\sum_{i=1}^{8}a_{i} \boldsymbol{u} _{i}^{\dagger})(x,+1,z) = 0  \vspace{0.5em}
\\
\quad\nabla\cdot(\overline{\boldsymbol u}^{n+1}+\sum_{i=1}^{8}a_{i} \boldsymbol{u} _{i}^{\dagger})(x,-1,z) = 0
\end{cases}\,.
\end{equation}

\section{Methods validation}
\label{sec:validation}
\begin{figure}
\centering
\includegraphics[width=0.99\linewidth]{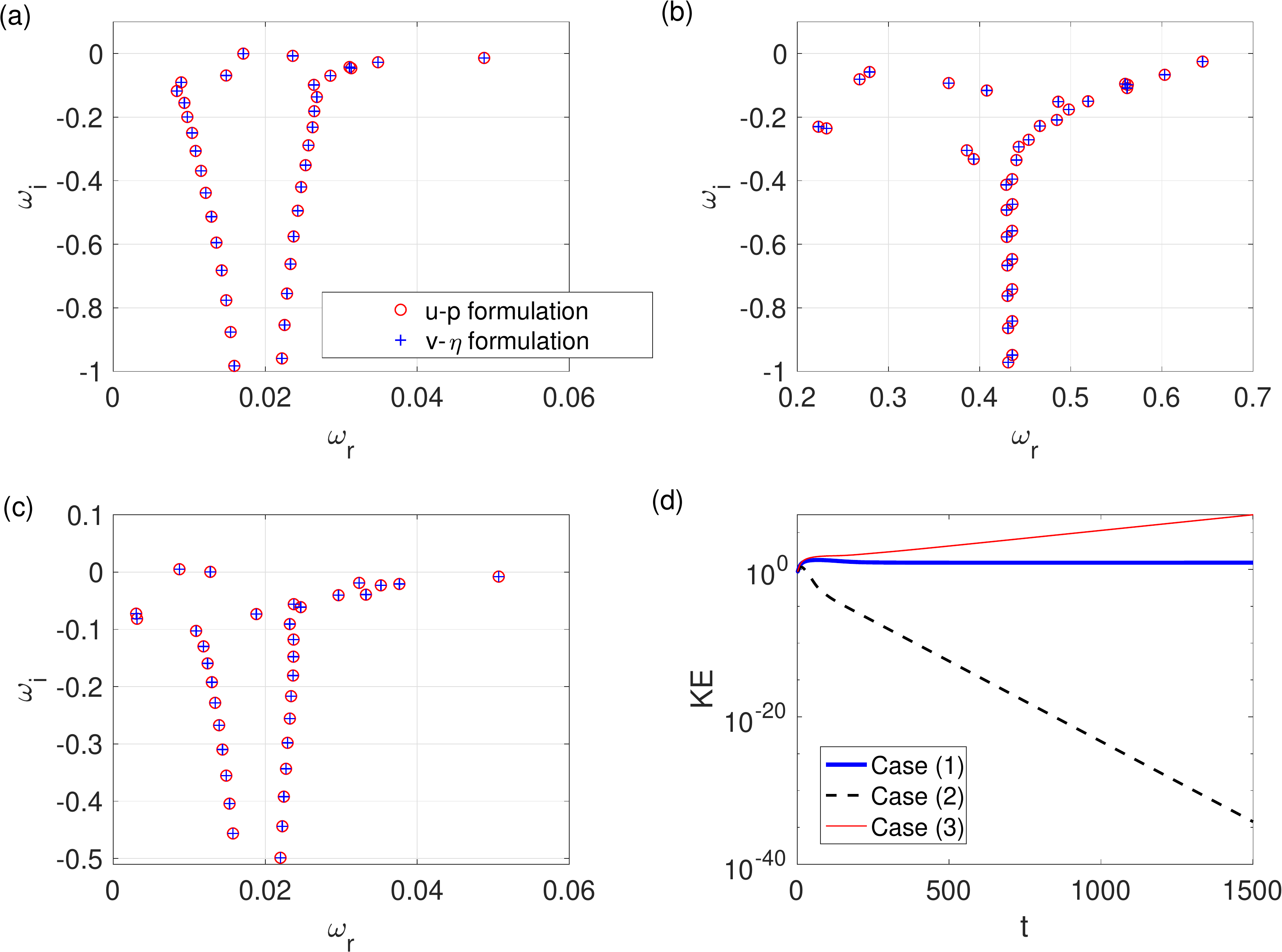}
\caption[validation]{
    \label{fig:validation} 
     (a-c) The eigenspectra calculated using the $\bm u-p$ formulation (circles) and $v-\eta$ formulation (crosses). The real and imaginary parts of the eigenvalues are denoted as $\omega_r$ and $\omega_i$, respectively. The parameters are detailed in table \ref{tab:validation}. Panels (a-c) correspond to Cases $(1)-(3)$, respectively.
     In (d), using the DNS formulation, the time-series of the kinetic energy ($KE$) of small perturbations in the three cases are shown. \textcolor{black}{The $KE$ is normalized by its value at $t=0$.} The exponential growth/decay rate can be calculated by Eqs.~\eqref{equ:growth_rate}.
} 
\end{figure}

Firstly, we validate our methods by calculating the eigenvalues using three different formulations described above. We consider the two-SH-wall channel with three sets of parameters (see the details of the parameters in table \ref{tab:validation}). 
The parameters are chosen such that Case (1) is nearly neutrally stable, Case (2) stable and Case (3) unstable. The eigenspectra are shown in figure \ref{fig:validation} and the leading eigenvalue $\omega_{\mathrm{max}}$ is given in table \ref{tab:validation}. For all these calculations, 128 Chebyshev grid points are used in the wall-normal direction. It can be seen that the eigenspectra calculated using our $\bm u-p$ and $v-\eta$ formulations agree well with each other.

In the DNS formulation, the simulations are performed with 128 wall-normal Chebyshev grid points and a time-step size of $\Delta t=0.005$. Figure \ref{fig:validation}(d) shows the time-series of the modal kinetic energy of small perturbations from the DNS formulation. From the time series, the decay/growth rates (equivalent to the imaginary part of the leading eigenvalue $\omega_i$) are calculated as
\begin{equation}\label{equ:growth_rate}
\gamma=\frac{1}{2}\frac{\log{KE(t_2)}-\log{KE(t_1)}}{t_2-t_1},
\end{equation}
where $\gamma$ denotes the growth rate, $KE=\int_V\bm u^2\text{d}V$ is the kinetic energy of disturbances, and $t_1$ and $t_2$ are two time instances 
in the exponential stage. 
Table \ref{tab:validation} shows that the growth rate calculated by the DNS formulation is very close to those by
the $\bm u-p$ and $v-\eta$ formulations. In summary, the excellent agreement between the three different formulations serves as a convincing validation of our eigenvalue calculation.

\begin{table}
\renewcommand\arraystretch{1.0}
\centering
\begin{tabular}{|c|c|c|}
\hline
parameters & formulation & $\omega_{\mathrm{max}}=\omega_r + i\omega_i $ \\
\hline
\multirow{3}{*}{Case (1): $Re=781$, $\alpha=0.08$, $\beta=-1.76$}
   & $u-p$       & 0.017110308575336 + 0.000001843885042i \\
   & $v-\eta$    & 0.017110308575723 + 0.000001843884918i \\
   & DNS         & $\gamma\approx 0.000001843946995$          \\
   \hline
\multirow{3}{*}{Case (2): $Re=781$, $\alpha=0.5$, $\beta=-1.76$}
   & $u-p$       & 0.644618135325376 $-$ 0.025061312033392i \\
   & $v-\eta$    & 0.644618135325378 $-$ 0.025061312033401i \\
   & DNS         & $\gamma\approx -0.025062686949318$                \\
   \hline
\multirow{3}{*}{Case (3): $Re=1500$, $\alpha=0.08$, $\beta=-1.76$}
   & $u-p$       & 0.008702345386942 $+$ 0.004892191113713i \\
   & $v-\eta$    & 0.008702345387000 $+$ 0.004892191113705i \\
   & DNS         & $\gamma\approx 0.004892128712797$                \\
\hline
\end{tabular}
\caption{\label{tab:validation}{Validation of the methods for the eigenvalue calculation. For all three test cases, we set $\lambda^\parallel=0.05$, $\lambda^\bot=0.025$ and $\theta=\pi/4$. The eigenvalue with the largest imaginary part, $\omega_{\mathrm{max}}$, is listed here, and the eigenspectra are shown in figure \ref{fig:validation}(a-c). For the DNS formulation, we only calculate the growth/decay rate of small disturbances according to Eqs.~\eqref{equ:growth_rate} using the time-series of the modal kinetic energy (see figure \ref{fig:validation}d).}}
\end{table}

\begin{figure}
\centering
\includegraphics[width=0.6\linewidth]{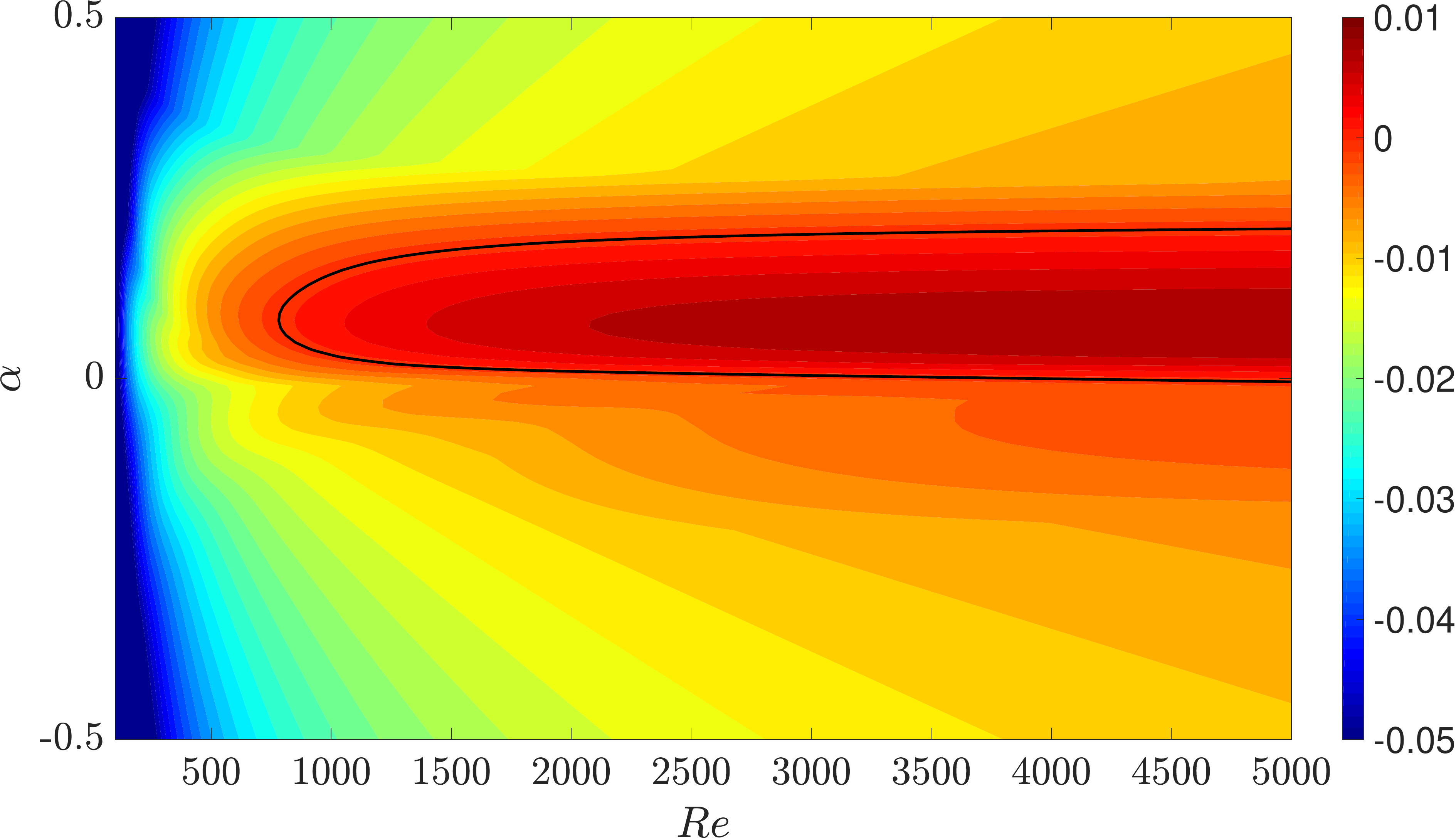}
\caption[alpha-Re-plane]{
    \label{fig:alpha_Re_plane} 
     The neutral curve (the bold line) in the $\alpha-Re$ plane for $\beta=-1.76$. The slip lengths are $\lambda^\parallel=0.05$ and $\lambda^\bot=0.025$ with $\theta=\pi/4$. The eigenvalue $\omega_i$ is plotted as the colormap.
} 
\end{figure}

\textcolor{black}{Besides, the grid convergence test for the Case (2) is performed by halving and doubling the grid number in the wall-normal direction. The results are shown in table~\ref{tab:convergence}. It is seen that, at least for calculating the leading eigenvalue, 64 Chebyshev grid points are already sufficient for all three formulations.}
\begin{sidewaystable}
\renewcommand\arraystretch{1.0}
\centering
\begin{tabular}{|c|c|c|c|}
\hline
 Case (2) & $ N=64$ &  $ N=128$ & $ N=256$ \\
\hline
    $u-p$       & 0.644618135325379 - 0.025061312033393i & 0.644618135325376 - 0.025061312033393i & 0.644618135325380 - 0.025061312033398i\\
    $v-\eta$    & 0.644618135325376 - 0.025061312033392i & 0.644618135325372 - 0.025061312033384i & 0.644618135325384 - 0.025061312033370i \\
    DNS         & $\gamma\approx  -0.025061672210072$  & $\gamma\approx    -0.025061672210055$  & $\gamma\approx   -0.025061672210053$      \\
   \hline
\end{tabular}
\caption{\label{tab:convergence}{The grid resolution convergence test for Case (2) as shown in table~\ref{tab:validation}. Grid numbers $N=64$, 128 and 256 are considered for this test.}}
\end{sidewaystable}

We note that the flow becomes linearly unstable already at around $Re=781$ (the Case 1) given $\lambda^\parallel=0.05$, $\lambda^\bot=0.025$ and $\theta=\pi/4$. \textcolor{black}{The neutral curve in the $\alpha-Re$ plane for $\beta=-1.76$ is shown in figure~\ref{fig:alpha_Re_plane}. In fact, the Case (1) is nearly at the nose of the neutral curve, i.e. $Re=781$ is nearly the critical Reynolds number at this slip setting.} Ref. \cite{Pralits2017} considered the same slip length setting whereas reported a critical Reynolds number around 9000 (see their FIG. 7), which according to our calculation was much overestimated.

\end{document}